\documentclass[manuscript]{aastex}

\usepackage{lscape}
\usepackage{multirow}
\usepackage{color}
\usepackage{ulem} 
\usepackage{hyperref}
\usepackage{ulem}       %
\makeatletter

\newcommand{\Rmnum}[1]{\expandafter\@slowromancap\romannumeral #1@}
\def\d    {\ifmmode {{\rlap{.}}^\circ}\else {${\rlap{.}}^\circ$}\fi}
\newcommand{\RNum}[1]{\uppercase\expandafter{\romannumeral #1\relax}}
\begin{document}
\title{Trigonometric Parallaxes of Star Forming Regions Beyond the Tangent
Point of the Sagittarius Spiral Arm}

\author{
Y. W. Wu\altaffilmark{1,2},
M. J. Reid\altaffilmark{3},
N. Sakai\altaffilmark{2},
T. M. Dame\altaffilmark{3},
K. M. Menten\altaffilmark{4},
A. Brunthaler\altaffilmark{4},
Y. Xu\altaffilmark{5},
J.J. Li\altaffilmark{5},
B. Ho\altaffilmark{5},
B. Zhang\altaffilmark{6},
K. L. J. Rygl\altaffilmark{7},
X.W. Zheng \altaffilmark{8}
}
\altaffiltext{1}{National Time Service Center, Key Laboratory of Precise
Positioning and Timing Technology, Chinese Academy of Sciences, Xi'an 710600,
China} 
\altaffiltext{2}{National Astronomical Observatory of Japan, 2-21-1 Osawa,
Mitaka, Tokyo 181-8588, Japan} 
\altaffiltext{3}{Center for Astrophysics $\vert$ Harvard \& Smithsonian, 
60 Garden Street, Cambridge, MA 02138, USA}
\altaffiltext{4}{Max-Planck-Institut f\"{u}r Radioastronomie, Auf dem H\"{u}gel
69, 53121 Bonn, Germany}
\altaffiltext{5}{Purple Mountain Observatory, Chinese Academy of Sciences,
Nanjing 210008, China}
\altaffiltext{6}{Shanghai Astronomical Observatory, Chinese Academy of
Sciences, Shanghai 200030, China}
\altaffiltext{7}{Italian ALMA Regional Centre, INAF$-$Istituto di
Radioastronomia, Via P. Gobetti 101, I-40129 Bologna, Italy}
\altaffiltext{8}{School of Astronomy and Space Sciences of Nanjing University,
Nanjing 210093, China}
\begin{abstract}
As part of the BeSSeL Survey, we report trigonometric parallaxes and proper
motions of molecular maser sources associated with 13 distant high mass star
forming regions in the Sagittarius spiral arm of the Milky Way.  
In particular, we obtain improved parallax distance estimates for three well 
studied regions: 
$1.9^{+0.1}_{-0.1}$ kpc for M17, 
$5.3^{+1.3}_{-0.9}$ kpc for W51, and 
$7.9^{+0.9}_{-0.7}$ kpc for GAL 045.5+00.0.
Peculiar motions for all but one source are less than 20 km~s$^{-1}$.  We fit a
log-periodic spiral to the locations and estimate an average pitch angle of $7\fdg2
\pm 1\fdg9$.  We find that the section of the arm beyond the tangent point in
the first quadrant of the Milky Way appears 15~pc below the IAU-defined
Galactic plane.
\end{abstract}


\section{Introduction}

Trigonometric parallaxes for molecular masers associated with high mass star
forming regions (HMSFRs) from the Bar and Spiral Structure Legacy (BeSSeL)
Survey and the VLBI Exploration of Radio Astrometry (VERA) project are
accurately tracing the spiral structure of the Milky Way.  As of 2015, over 100
parallaxes had been published \citep{2014ApJ...783..130R,  honma2015}.  These
contained 18 parallaxes for sources in the Sagittarius spiral arm, most of
which were located closer to the Sun than the arm's tangent point at
$l\approx$49$^\circ$ and a distance of $\approx5$~kpc
\citep[]{2014AA...566A..17W}.  Beyond the tangent point, only four parallax
measurements had been made, and those tended to have large uncertainties ($>$1
kpc).  With the aim of better constraining the
structure of the ``far'' side of the Sagittarius arm (Sgr Far, i.e., beyond the
tangent point), we measured
parallaxes of 13 additional maser sources.  In Section 2 we describe the
observations and data analysis.  Parallaxes and proper motions are presented in
Section 3.  We discuss how sources were assigned to the Sgr Far arm in Section 4.
The geometry and kinematics of the arm are discussed in Section 5
and 6, and we summarize our conclusions in Section 7.

\section{Observations and Data Analysis} \label{sect-observations}

Between 2012 and 2015, we conducted parallax observations of eleven 6.7-GHz
methanol and two 22-GHz water masers in the Sag Far arm segment using the
National Radio Astronomy Observatory's\footnote{The National Radio Astronomy
Observatory is a facility of the National Science Foundation operated under
cooperative agreement by Associated Universities, Inc.} Very Long Baseline
Array (VLBA).  Generally, three masers nearby on the sky were observed within
the same seven-hour track.  The dates of the observations are listed in Table
\ref{tbl-A1}, and  coordinates of masers and background quasars are presented
in Table \ref{tbl-A2}.   

The observations used the same equipment setup and calibration procedures
described by \citet{2016SciA....2E0878X}.  The data were correlated with the
DiFX\footnote{DiFX, a software correlator for VLBI, is developed as part of the
Australian Major National Research Facilities Programme by the Swinburne
University of Technology and operated under licence.} software correlator
\citep{2007PASP..119..318D} at the VLBA correlation facility in Socorro, NM.
Continuum emissions were recorded with four adjacent dual-circular polarization
intermediate frequency (IF) bands of 16 MHz and correlated with
32 channels per IF. The maser line emissions were set at band center of the
second or third IF and were correlated with 2000 and 4000 spectral channels for
BR149 and BR198 observations, respectively, yielding velocity channel spacings
of 0.36 and 0.18 km~s$^{-1}$ for the 6.7-GHz masers and 0.11 km~s$^{-1}$ for
the 22 GHz masers.

Generally, the channel with the strongest maser emission was selected as the
interferometer phase reference used to calibrate both the maser and background
quasar data.  For G041.15$-$00.20 and G041.22$-$00.19, whose background quasars
(J1907+0907, J1919+0619) were strong enough to fringe-fit delays with precision
better than 100 ps, we performed an additional calibration to remove residual
(assumed dispersive) delays before phase referencing.  After calibration, images of
continuum emission of the quasars and cubes of maser-line emission were
produced with AIPS task IMAGR.   The task JMFIT was used to fit two-dimensional
Gaussian brightness distributions to compact emission peaks, yielding positions
of maser spots and background quasars. 

\section{Parallax Estimation}\label{sect-parallax}

For 22-GHz water masers, fitting parallax and proper motion of the masers
(relative to background quasars) is straightforward and has been discussed in
detail in \cite{Reid2009a}.  However, at the lower observing frequency of the
6.7 GHz methanol masers, the effects of unmodeled ionospheric ``wedges'' above
individual antennas can result in systematic position shifts across the sky
(relative to different quasars), complicating the parallax estimation.
\citet{2017AJ....154...63R} discuss this problem in detail and present a
strategy to minimize the effects on parallax estimates.   This involves a
two-step method of, first, using the relative position data for multiple
quasars to generate an ``artificial quasar'' at the location of the target maser
and, second, fitting parallax and proper motion parameters to the artificial
quasar data.  While this technique improves parallax accuracy, compared to
averaging parallax results for each quasar, with our typical 4-epoch data it
yielded poor estimates of the parallax {\it uncertainty}.  

Since the publication of \citet{2017AJ....154...63R}, we developed a better
approach that preserves the accuracy of the artificial quasar method, while
improving the {\it uncertainty} estimates.  In a single step, we modeled the
positional data of a maser spot relative to multiple quasars at epoch $t$ as
the sum of the maser's parallax and proper motion and a planar ``tilt,'' owing
to ionospheric wedges, of the quasar positions about the maser position: 
\begin{equation}
\Delta \theta _{s,q}^{x}(t) = \Pi ^{x} (t)+(\Delta \theta _{s}^{x}-\Delta \theta _{q}^{x})+\mu _{s}^{x}\delta t + S_{x}^{x}(t)\Theta _{q}^{x}+ S_{y}^{x}(t)\Theta _{q}^{y}
\label{eq-1}
\end{equation}
In Eq.~\ref{eq-1}, $\Pi ^{x} (t)$ is the $x$-component of the parallax shift;
$\Delta\theta_{s}^{x}$ and $\Delta\theta _{q}^{x}$ are constant offsets of
maser spot, $s$, and QSO, $q$, from the position used in correlation; $\mu
_{s}^{x}\delta t$ is the $x$-component of maser spot position shifted by proper motion;
$S_{x}^{x}(t)\Theta _{q}^{x}$ is the $x$-position shift owing to an ionospheric
wedge (``slope'' in mas~deg$^{-1}$) in the $x$-direction times the
$x$-component of the separation between the maser and QSO $q$ (in deg);
$S_{y}^{x}(t)\Theta _{q}^{y}$ is the $x$-position shift owing to an ionospheric
wedge in the $y$-direction times the
$y$-component of the separation between the maser and QSO $q$ at epoch
$t$.  The offset of one maser spot was set to zero and held constant, since one
cannot solve for all $\Delta\theta^x$ terms with relative position information.
There is an analogous equation that holds for the $y$-coordinate. 

We used a Markov chain Monte Carlo approach to generate probability
distribution functions for all parameters, varying all parameters
simultaneously, and then fitted the marginalized distribution functions 
to obtain the parallaxes and
proper motion values given in Table \ref{tbl-1}.  Typical 6.7 GHz maser
parallax uncertainties are $\pm0.02$ mas, which correspond to $\pm1.3$ kpc
uncertainty at a distance of 8 kpc, which is characteristic of our sources past
the tangent point of the Sagittarius arm.

Regarding G052.10+1.04, \citet{Oh2010} reported a parallax of
0.251$\pm$0.036~mas measured at 22 GHz by the VERA project, whereas our 6.7-GHz
data yields a parallax of 0.162$\pm$0.013~mas. Formally, the parallax
difference has $2.3\sigma$ tension.  However, we note that \citet{Oh2010}
fitted the parallax for G052.10+1.04 by combining four maser spots, implicitly
assuming the position differences of these maser spots with respect to their
reference QSO are uncorrelated.  This gives an optimistic parallax uncertainty,
since atmospheric delay residuals between the maser and the QSO, which are
common to all maser spots, usually dominate systematic uncertainties.  BeSSeL
Survey results generally allow for correlated parallaxes among maser spots, as
we conservatively inflate the formal uncertainty by $\sqrt{N}$, where $N$ is
number of spots used for parallax fitting.  Doing this for the VERA parallax
would give an uncertainty of $\pm0.072$ mas, making these two parallaxes
statistically consistent at the $1.2\sigma$ level.  Then combining VERA and BeSSeL
measurements yields a parallax of 0.165$\pm$0.013~mas.

For G043.89$-$0.78, our 6.7-GHz data yields a parallax of 0.144$\pm$0.017~mas,
and combining this with the 22~GHz parallax of 0.121$\pm$0.020~mas of
\citet{2014AA...566A..17W} yields a variance-weighted average parallax of 
0.134$\pm$0.013~mas.
Rather than averaging the proper motions from the two maser
species, we adopt the motion of the 6.7 GHz methanol masers as the best
estimate of the exciting star's motion.  We do this because 22 GHz water
masers trace outflows with speeds of typically tens of km~s$^{-1}$,
whereas methanol masers are generally within 5~km~s$^{-1}$ of the exciting
star.

We note that in our sample, masers with similar
coordinates, distances and radial velocities are likely in the same
giant molecular cloud.   For example, G014.63$-$00.57 and G015.03$-$00.67
are associated with the large star forming region M17 (see Figure \ref{fig-A1});
G049.48$-$00.36, G049.48$-$00.38, G048.99$-$00.29 and G049.19$-$00.33 appear
associated with W51 (see Figure \ref{fig-A2}); and G045.45+00.06, G045.07+00.13
and G045.49+00.12 may be associated with the H\Rmnum{2} region GAL 045.5+00.0
(see Figure \ref{fig-A3}).  By combining parallax measurements of the
associated sources, we can improve the distance estimates to the larger
star forming regions, yielding averaged parallaxes of 0.524$\pm$0.024~mas for M17,
0.189$\pm$0.037~mas for W51, and 0.126$\pm$0.013~mas for GAL 045.5+00.0.

\section{Assigning Sources to Arm Segments}\label{sect:assignments}

\begin{figure}[ht]
\figurenum{2}
\centering
\includegraphics[width=14cm]{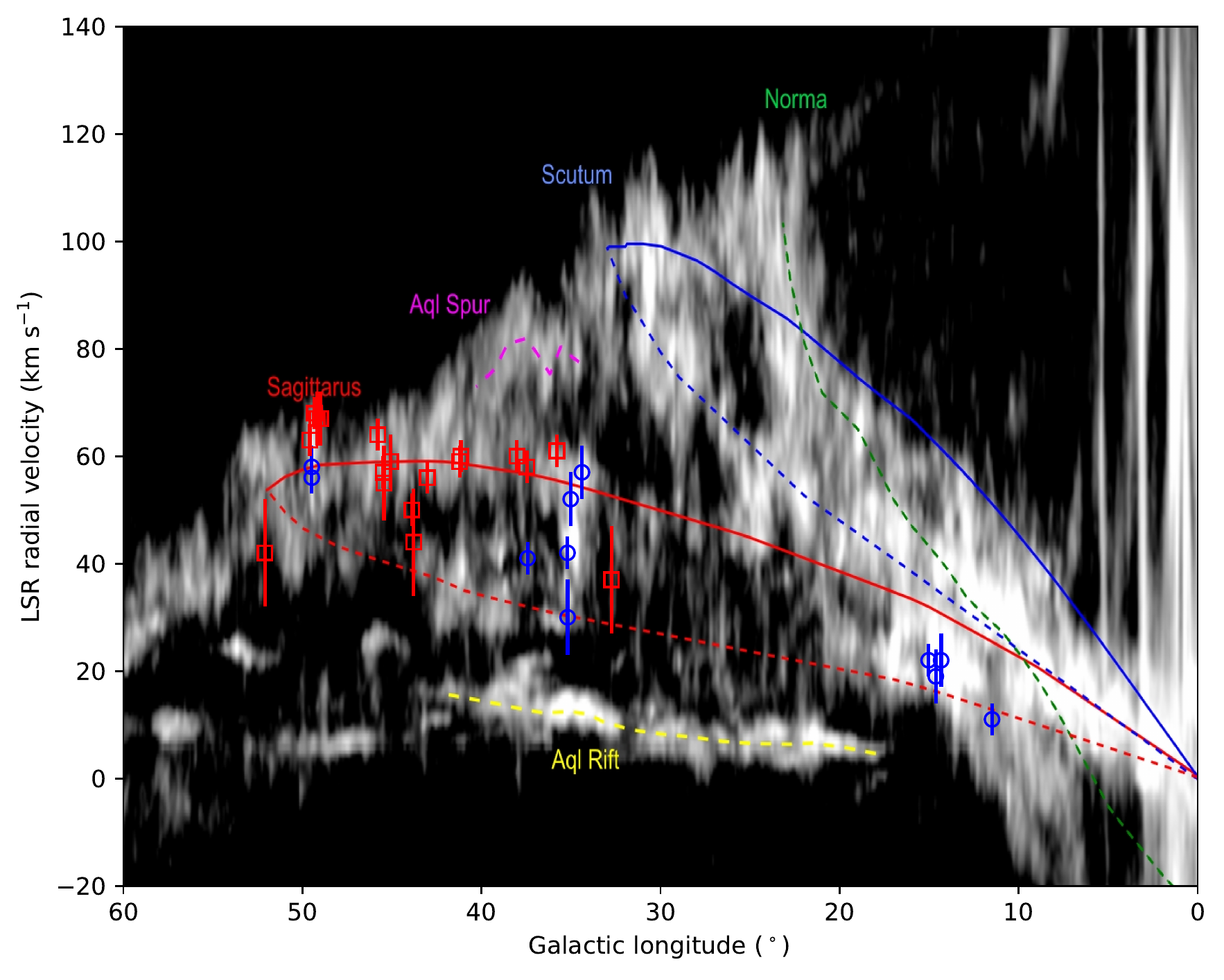}
\caption{Overlay of the Sgr Far (red squares) sources and Sgr Near (blue circles) 
on a longitude-velocity plot of CO emission from the CfA 1.2m survey \citep{Dame2001}. 
This figure is adapted from Figure~7 of \citet{2016ApJ...823...77R}, where the Sagittarius, 
Scutum, and Norma arms in the first Galactic quadrant, as well as the Aquila spur and 
Aquila Rift sub-structures are traced by colored lines. The near and far segments of the arms,
with respect to the arm tangent point, are indicated by {\it dashed} and {\it solid} lines, respectively.
Our assignment of sources to the near or far segments used this and other information 
(see Section \ref{sect:assignments} for discussion).
\label{fig-2}}
\end{figure}

In Figure \ref{fig-2}, we show the locations 
of HMSFRs with measured parallaxes superposed on a CO longitude-LSR velocity ($l$-$v$) 
plot.  The assignment of a maser source to a specific spiral arm or arm segment
can usually be done via such a plot, by comparing the Galactic coordinates and LSR velocity of the source with continuous traces of CO (or HI) emission
previously associated with an arm.  However, ($l$-$v$) information alone does
not always allow a definitive assignment.  Even for sources with direct parallax
distances, the distance uncertainty does not always guarantee a spiral arm
segment assignment.  Instead, distance can best be assigned on a probabilistic
basis, and when doing this it is important that all possible information is
utilized.
  
\begin{figure}[ht]
\figurenum{3}
\centering
\includegraphics[width=14cm]{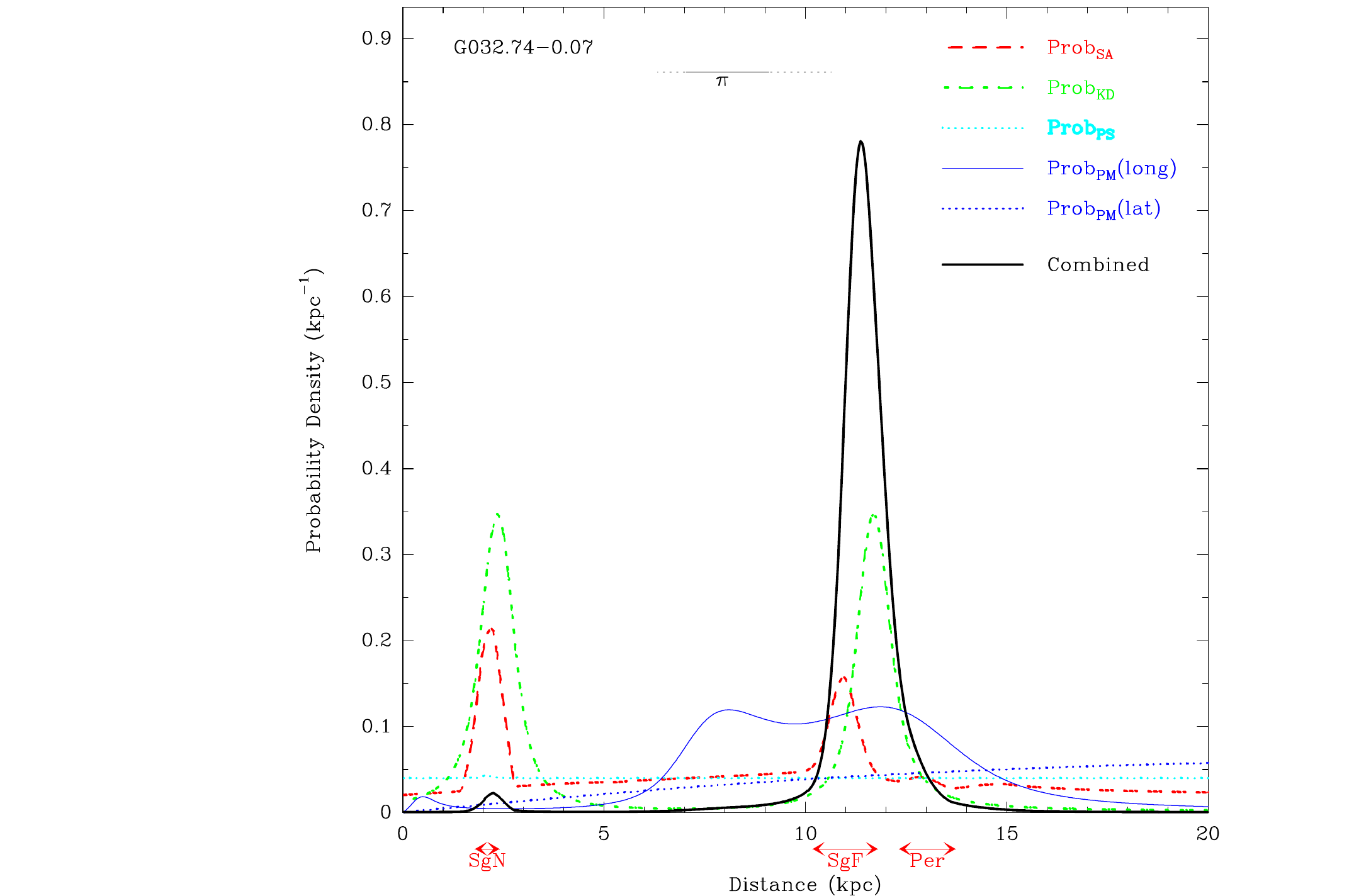}
\caption{Distance probability density function for G032.74$-$0.07, a 22 GHz
water maser at $V_{LSR}=47$ km~s$^{-1}$. The ($l, b, v$) values slightly favor
the Sgr Near (SgN) arm segment. However, the combination (heavy black line) of
arm assignment probability (red dashed line) with the radial-velocity kinematic distance (green
dotted-dashed line), the proper motion in the Galactic longitude and latitude
directions (blue solid and dotted lines) favors location in the Sgr Far (SgF)
arm segment over the SgN segment and Perseus arm (Per).  Our measured parallax
for this source is shown at the top of the figure by the horizontal solid and
dotted lines, indicating the $1\sigma$ and $2\sigma$ uncertainty ranges.
The parallax also favors the Sgr Far arm segment, albeit with $2\sigma$
tension.
\label{fig-3}}
\end{figure}

For the BeSSeL Survey, we evaluate arm segment assignment using our Bayesian
distance estimator \citep{2016ApJ...823...77R}, which considers all
distance-related information, including a new feature which incorporates a
proper motion ``kinematic-distance'' probability density function 
(PDF; Reid et al., in preparation). For example,
consider G032.74$-$00.07, a water maser at $v=37$ km~s$^{-1}$, which in ($l$,
$v$) space lies between the Sgr Near and Sgr Far arm segments\footnote{We define 
"near" and "far" for the Sgr arm as relative to the tangent point.  While this is
equivalent to kinematic near/far convention, it does not imply that we have
used absorption information to make this distinction.} (as seen in Figure
\ref{fig-2}). The distance PDF for this source is shown in Figure \ref{fig-3}.
As expected, the probabilities of matching a spiral arm segment, Prob$_{SA}$,
are similar for the Sgr Near and the Sgr Far segments.  However, the Galactic
longitude motion, Prob$_{\rm PM}$(long), favors the Sgr Far segment, as does
the measured parallax, denoted by the horizontal line at the top of Figure
\ref{fig-3} (with solid and dotted lines indicating its 1-sigma and 2-sigma
uncertainty ranges), albeit with about a 2-sigma deviation. Putting all
information together, we assigned G032.74$-$00.07 to the Sgr Far arm segment.
Based on similar evidence, we also assigned G043.79$-$0.12 to Sgr Far. 

Other apparent anomalies in the arm segment assignments are seen for the 
following sources near $l=35^\circ$ in Figure~\ref{fig-2}:  
G034.39+00.22, G035.02+00.34, and G035.20$-$01.73.
This is a very interesting region, with sources projected close to the W44 SNR.  
It is possible that the precursor O-star's wind 
may have accelerated gas that has now formed the stars we observe \citep{Reid1985}.   
If so, this might explain anomalous velocities which can confuse ($l,v$) locations.   
For sources in this region, we rely mostly on parallaxes to assign to an arm segment.

\section{Geometry of the Sagittarius arm}

\subsection{Galactic Locations of HMSFRs}

\begin{figure}[ht]
\figurenum{4}
\centering
\includegraphics[width=14cm]{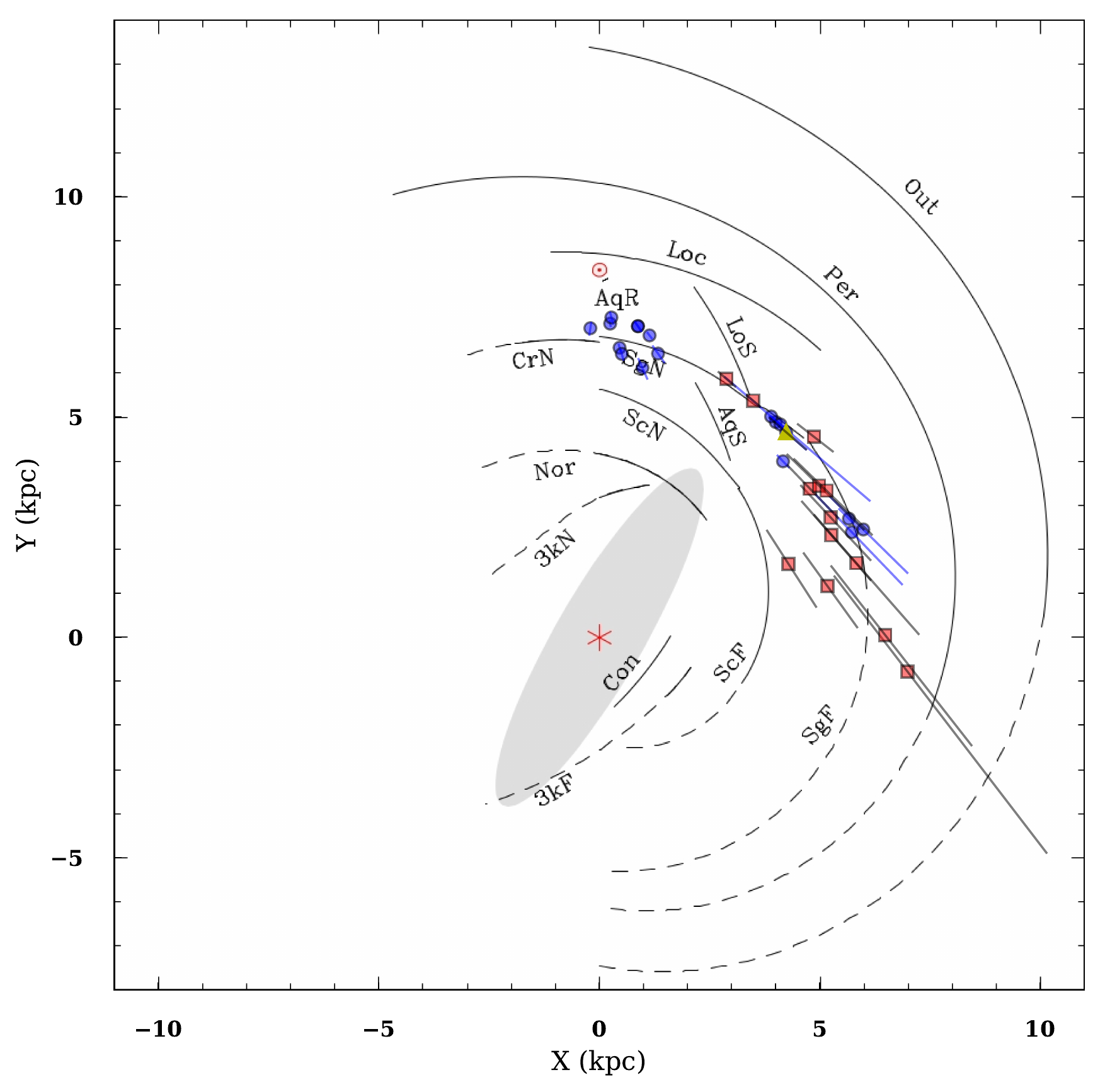}
\caption{Locations of HMSFRs in the Sagittarius arm on a schematic plan view of the 
Milky Way from \citet{2016ApJ...823...77R}. Red squares are 13 sources reported
in this paper; blue circles are 18 HMSFRs from \citet{2014AA...566A..17W}. The
yellow triangle is G048.99$-$00.29 from \citet{2015PASJ...67...65N}.
Sources are located by the parallax measurement, assuming distance is the
inverse of the parallax, and the lines indicate distance uncertainty by
adding and subtracting the parallax uncertainty from the parallax.
\label{fig-4}}
\end{figure}

In Figure \ref{fig-4}, we show locations of all HMSFRs associated with the Sgr arm
with measured parallaxes superposed on a plan view of the Galaxy. There now are
18 HMSFRs located beyond the tangent point of the Sgr arm: 13 sources reported
in this paper, 4 reported by \citet{2014AA...566A..17W}, and 1 reported by
\citet{2015PASJ...67...65N}.  The locations of all sources are consistent with
the \citet{2016ApJ...823...77R} model for the Sagittarius arm.

\begin{figure}[ht]
\figurenum{5}
\centering
\includegraphics[width=16cm]{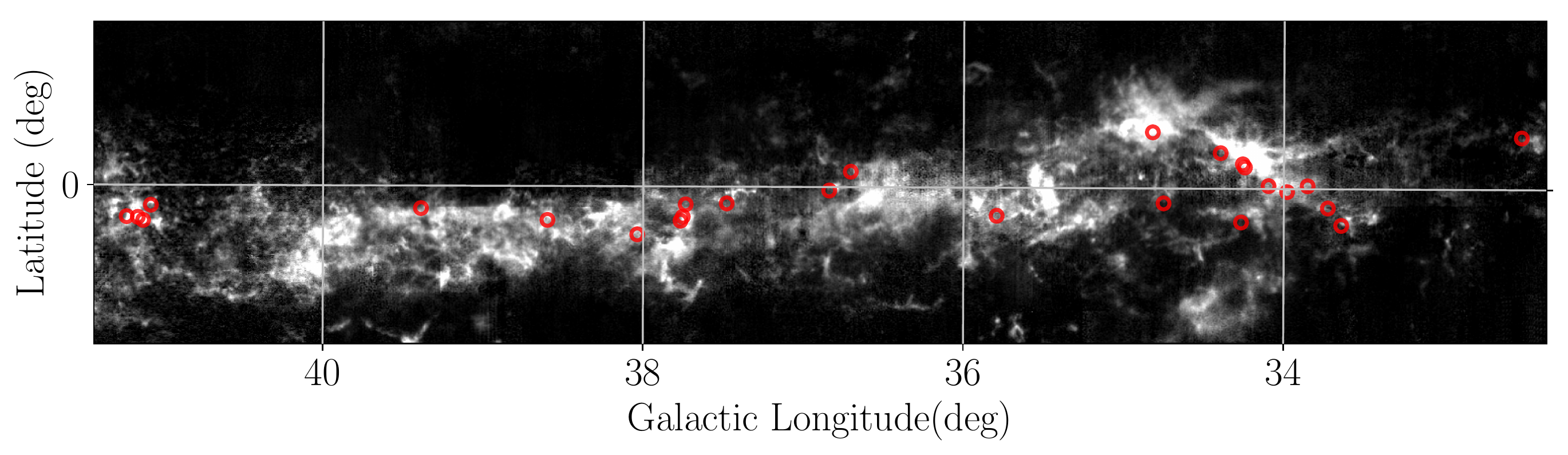}
\includegraphics[width=16cm]{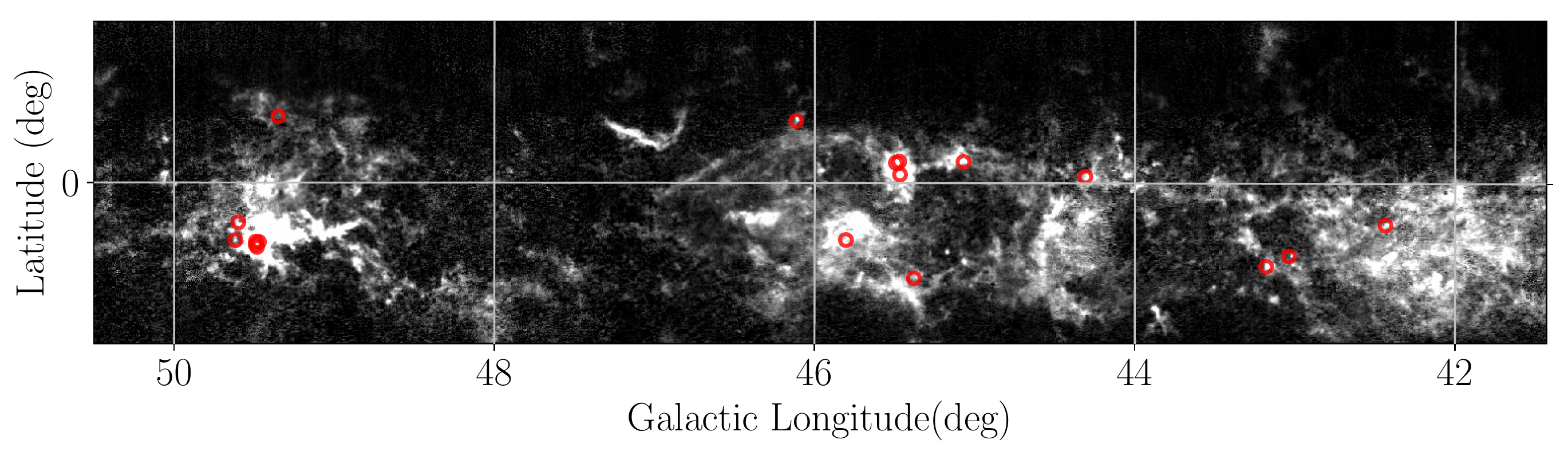}
\caption{Intensity map of the $^{13}$CO (J = 1-0) emission integrated over
50~$<$~V$_{LSR}$~$<$~70~km~s$^{-1}$ from the Galactic Ring
Survey \citep[][]{2006ApJS..163..145J}. Superposed with {\it red circles} are methanol masers
from the Parkes Methanol Multibeam Survey \citep{2015MNRAS.450.4109B} with
50~$<$~V$_{LSR}$~$<$~70~km~s$^{-1}$.  \label{fig-5}} 
\end{figure}

We note that 16 of 18 sources in the far portion of the Sagittarius arm are
below the IAU-defined Galactic plane.  In order to investigate this asymmetry,
we looked at the Galactic distribution of all the 6.7-GHz masers thought to be
in this section of the arm based on their location in the $l$-$v$ plane.   We
assigned sources to the far arm section if their longitudes were within the range
32$^\circ$$<l<$ 50$^\circ$ and their velocities were within the range
50~$<$~V$_{LSR}$~$<$~70~km~s$^{-1}$.  Searching the Methanol Multi-Beam Survey
maser catalog of \citet{2015MNRAS.450.4109B}, we identified 42 6.7-GHz masers
in the far portion of the arm; they are shown as red circles in Figure
\ref{fig-5}. 

\begin{figure}[ht]
\figurenum{6}
\centering
\includegraphics[width=12cm]{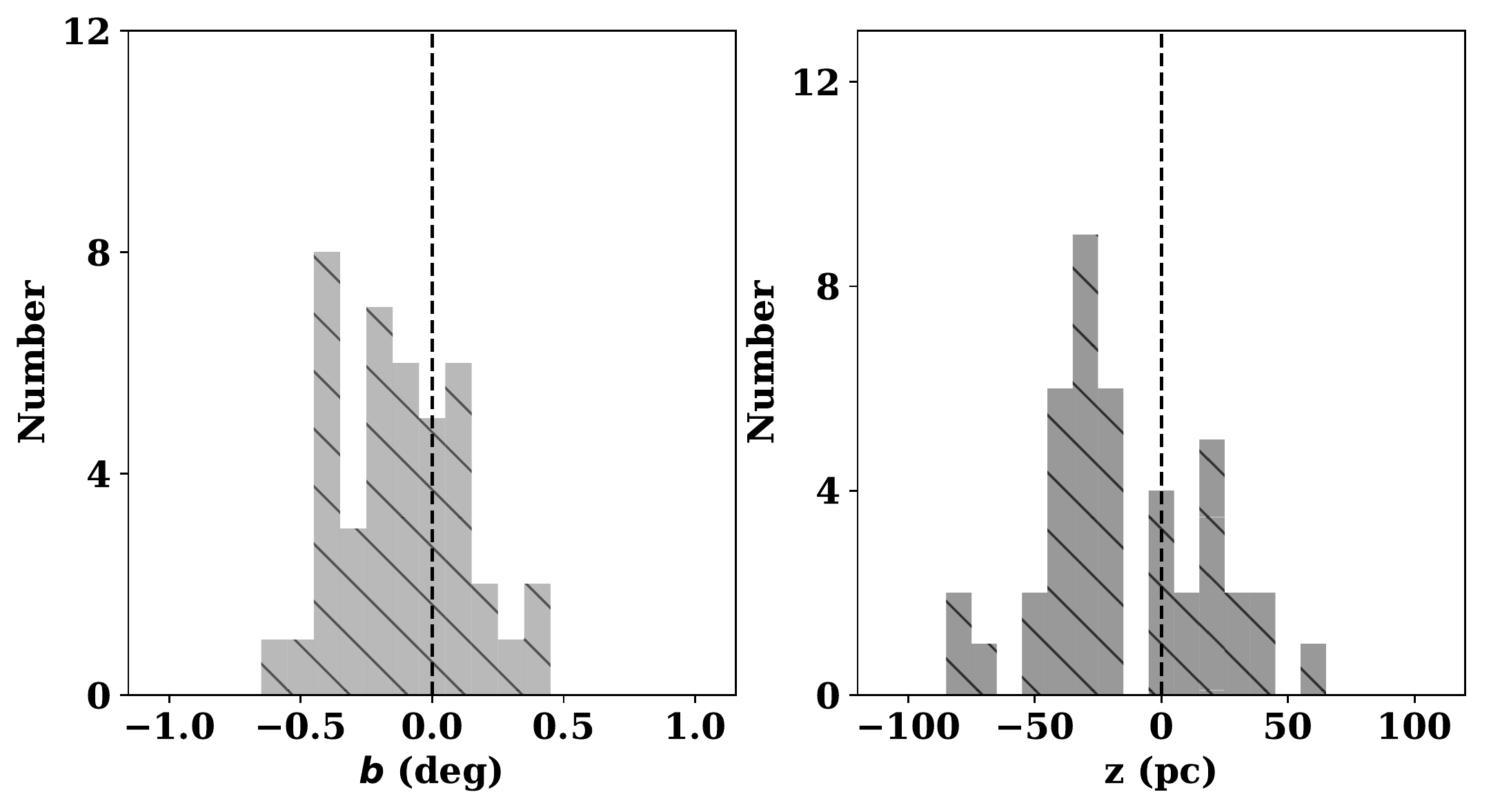}
\caption{Histograms of the Galactic latitudes and z-heights 
of 6.7-GHz methanol masers from the methanol multi-beam survey
\citep{2015MNRAS.450.4109B} with 32$^\circ$$<l<$ 50$^\circ$ and 50~$<V_{\rm
LSR}<$~70 km s$^{-1}$. 
\label{fig-6}}
\end{figure}

In the two panels of Figure \ref{fig-6} we show the histograms of Galactic
latitude, $b$, and the vertical distances, $z$, of these masers from the IAU
Galactic plane.  For sources without maser parallaxes, we used distances
calculated with the Bayesian distance estimator of \citet{2016ApJ...823...77R}.
Figure \ref{fig-6} suggests a systematic z-offset for the Sgr Far arm section.
The number of sources below/above the Galactic mid-plane are 28/14
respectively; the mean vertical distances with respect to the mid-plane, $<z>$,
of these 6.7-GHz methanol masers is $-15$~pc.

\subsection{Pitch Angle of the Sagittarius Arm}

\begin{figure}[ht]
\figurenum{7}
\centering
\includegraphics[width=14cm]{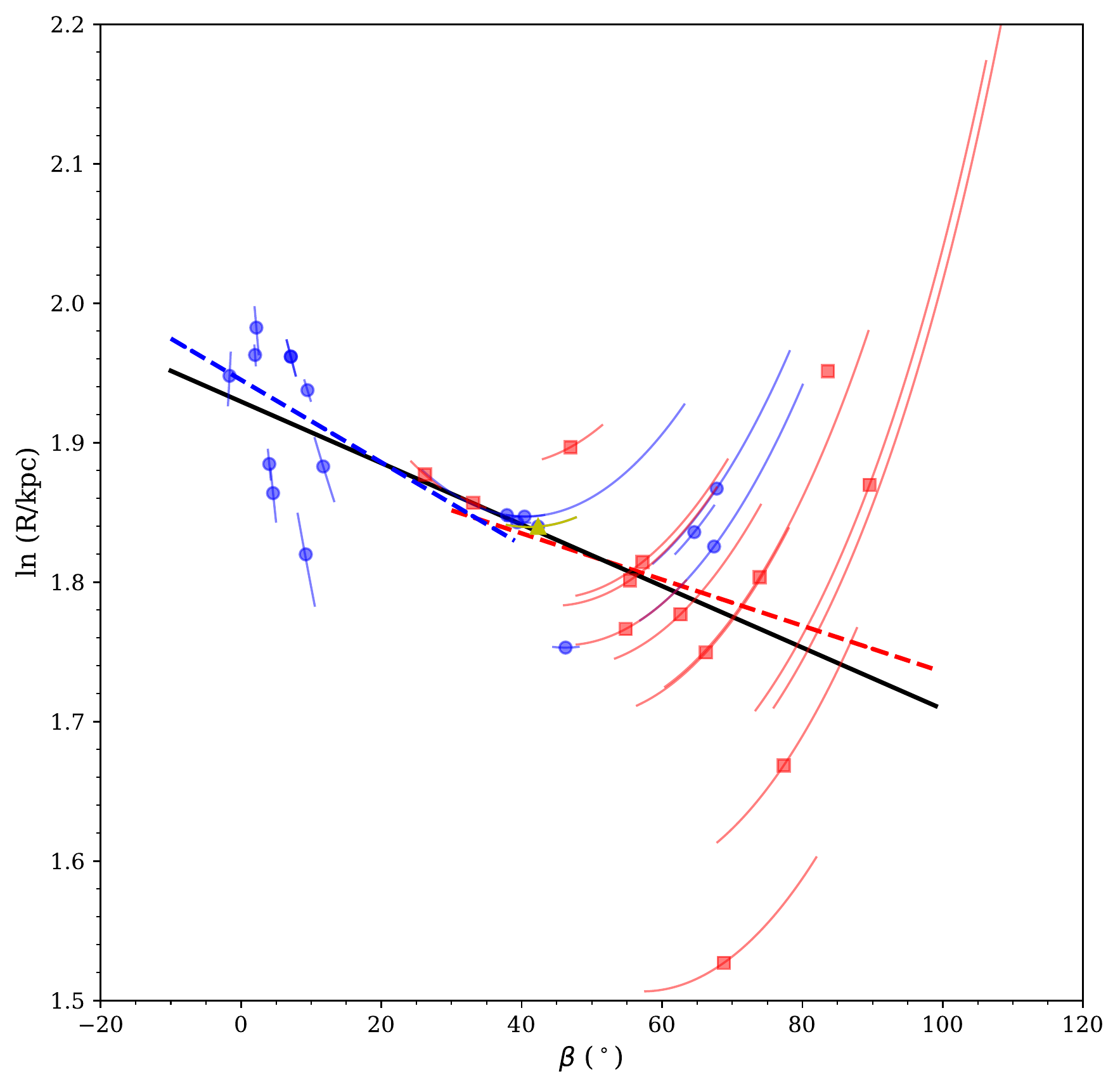}
\caption{Logarithm of Galactic radius (in kpc units), R, versus Galactocentric azimuth, $\beta$. Symbols are the same as in
Figure \ref{fig-4}. The black solid line denotes a pitch angle of 7$\fdg$2,
fitted with all points; the red and blue dashed lines denote pitch angles of
5$\fdg$4 and 9$\fdg$4 fitted to the Sgr Far and Near arm segments, respectively.
\label{fig-7}}
\end{figure}

A spiral arm pitch angle, $\psi$, can be defined as the angle between an arm
segment and a tangent to a Galactocentric circle.  For a log-periodic spiral,
$\ln(R) = \ln(R_{ref}) - (\beta-\beta_{ref}) \tan\psi$, where $R$ is the
Galactocentric radius in kpc at azimuth $\beta$ in radians; 
$\beta_{ref}$ is a reference azimuth and $R_{ref}$ is the radius of the arm at 
that azimuth.  
With parallaxes for 30 high mass star forming regions, we estimate a
global pitch angle of 7$\fdg$2~$\pm$~1$\fdg$9, which is consistent with our previous
estimate of (7$\fdg$1~$\pm$~1$\fdg$5) \citep{2014AA...566A..17W}.  The increase in
uncertainty between the previous and current estimates suggests that the pitch
angle may not be constant over the larger range of azimuth that we now have
sampled.  Such effects are seen in other spiral galaxies
\citep{2015ApJ...800...53H}.  If we divide the arm into two sections, we
estimate pitch angles of 5$\fdg$4~$\pm$~3$\fdg$3 and 9$\fdg$4~$\pm$~3$\fdg$9 for
the far and near portions of the Sgr arm, respectively, as shown in Figure \ref{fig-7}.  
For comparison, pitch angles of spiral arms in the Milky Way, determined by 
maser parallaxes range from $7^\circ$ to $20^\circ$, with the Sagittarius arm 
having the smallest pitch angle \citep{2014ApJ...783..130R}.

\section{Kinematics of the Sagittarius arm}

We now investigate the 3-dimensional motions of HMSFRs in the Sgr arm.  We
follow the methodology given in the Appendix of \citet{Reid2009b} to calculate
peculiar motion: (U$_s$, V$_s$, W$_s$), where U$_s$ is toward the Galactic center,
V$_s$ is in the direction of Galactic rotation, and W$_s$ is toward the North Galactic Pole. We adopt the ``Universal'' form for
the rotation curve of \citet{1996MNRAS.281...27P}, with a circular rotation
speed at the Sun of $\Theta_0$=241 km s$^{-1}$, R$_0$=8.31 kpc, and Solar
Motion components of (U$_\odot$, V$_\odot$, W$_\odot$) = (10.5, 14.4, 8.9) km
s$^{-1}$ (see the ``Univ'' model in Table 5 of \citet{2014ApJ...783..130R}).

\begin{figure}[ht]
\figurenum{8} \centering
\includegraphics[width=16cm]{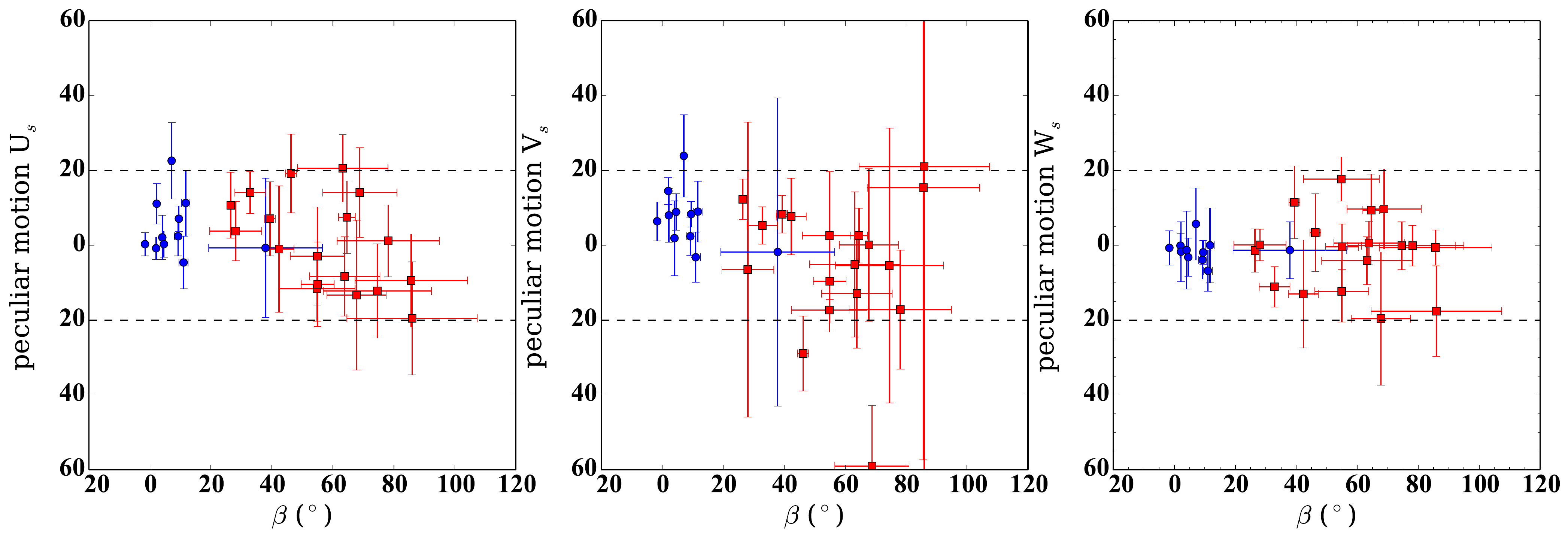} \caption{Peculiar (non-circular)
motion components in km~s$^{-1}$ of the Sgr arm HMSFRs as a function of the Galactocentric
azimuth, $\beta$. Blue circles and red squares denote sources on the near and far side 
of the tangent point, respectively. Except for one source, G032.74$-$00.07, the
peculiar motions of all
HMSFRs are within about $\pm$20~km~s$^{-1}$ (denoted by dashed lines). \label{fig-8}}
\end{figure}

In Table \ref{tbl-2}, we list peculiar motion components for the 30 (18 Sgr Far
and 12 Sgr Near) HMSFRs in the Sgr arm with measured parallaxes and proper
motions.   Plots of the peculiar motion components versus Galactocentric
azimuth $\beta$ are shown in Figure \ref{fig-8}.  All values are consistent
with $0 \pm 20$~km~s$^{-1}$, except for the $V$ component of G032.74$-$00.07,
which is $-59\pm16$~km~s$^{-1}$.  The variance-weighted average peculiar motion
components for the far, near and all Sgr arm sources are ($4.3\pm2.4$, $-0.6\pm2.2$,
$-0.2\pm1.6$)~km~s$^{-1}$, ($2.0\pm1.4$, $7.1\pm1.5$,
$-1.1\pm1.5$)~km~s$^{-1}$ and ($2.6\pm1.1$, $4.5\pm1.3$,
$-0.7\pm1.1$)~km~s$^{-1}$, repectively.

\section{Conclusions} \label{sect-conclusions}

We have measured trigonometric parallaxes and proper motions for 13 HMSFRs, which
are located on the far side (past the tangent point) of the Sagittarius arm.
Together with 5 published maser parallaxes, we studied the location, shape and
kinematics of the far portion of the Sgr arm.  We find that the 6.7-GHz masers
in this arm section are on average $15$~pc below the IAU-defined Galactic
mid-plane. The average pitch angle of the Sgr arm, now measured over $\approx10$ kpc is
$7\fdg2 \pm 1\fdg9$, but there is evidence that the pitch angle may not be constant along
the arm. Peculiar motions for all but one source are less than 20 km~s$^{-1}$, but
the there is a small average motion toward the Galactic center and in the direction of
Galactic rotation.

\vspace{5mm}
\clearpage

\acknowledgements This work was supported by the Chinese NSF through grants NSF
11673066, NSF 11673051, NSF 11873019. B. Zhang and Y. W. Wu are supported by
the 100 Talents Project of the Chinese Academy of Sciences (CAS). K. L. J. Rygl
acknowledges financial support by the Italian Ministero dell'Istruzione
Universit\`{a} e Ricerca through the grant Progetti Premiali 2012-iALMA (CUP
C52I13000140001).  Y. W. Wu is supported in part by the West Light Foundation
of The Chinese Academy of Sciences (XAB2016A06). 

{\em Facilities:} VLBA

\begin{figure}[ht]
\figurenum{1}
\centering
\includegraphics[height=6.5cm]{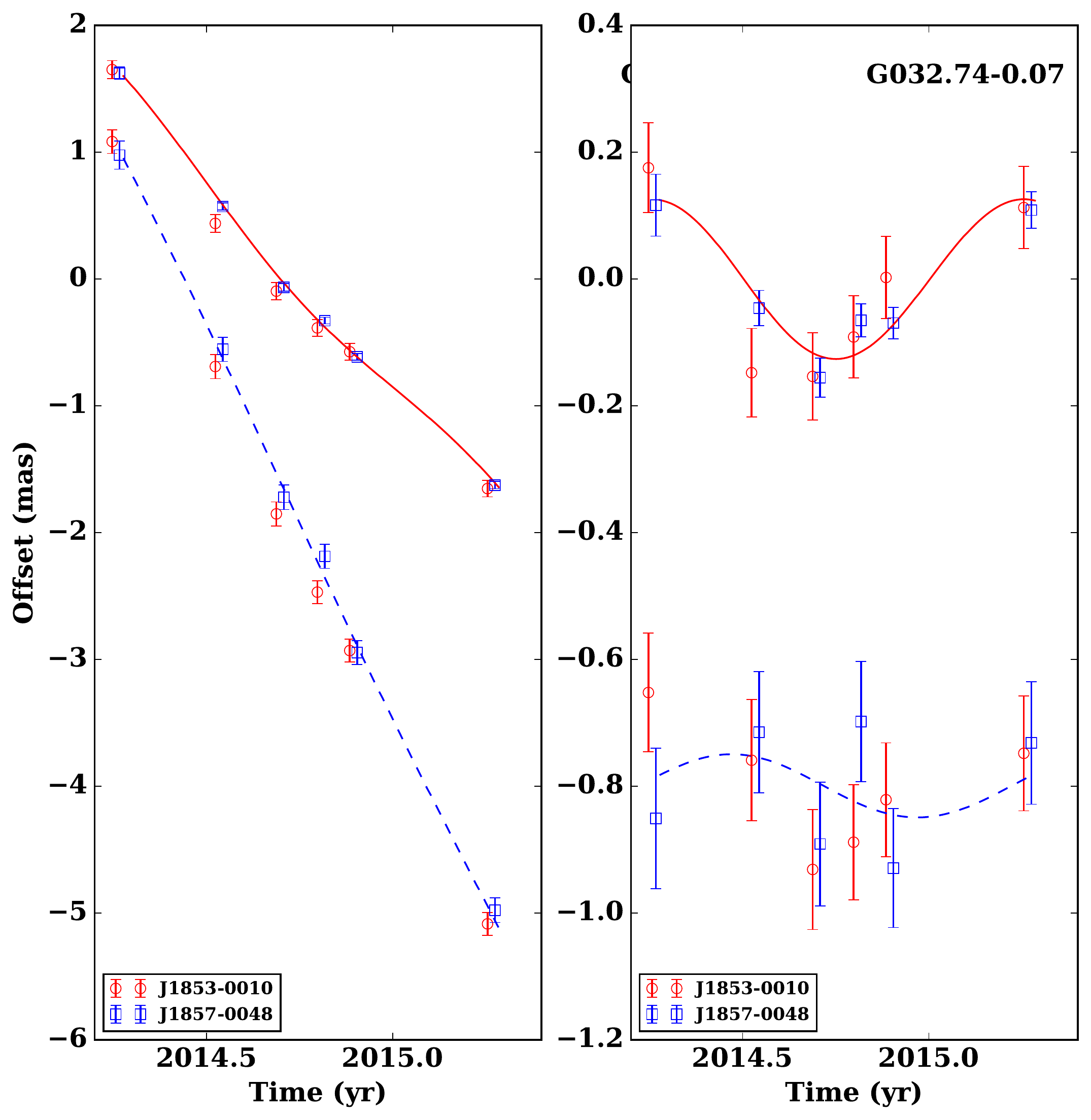}
\includegraphics[height=6.5cm]{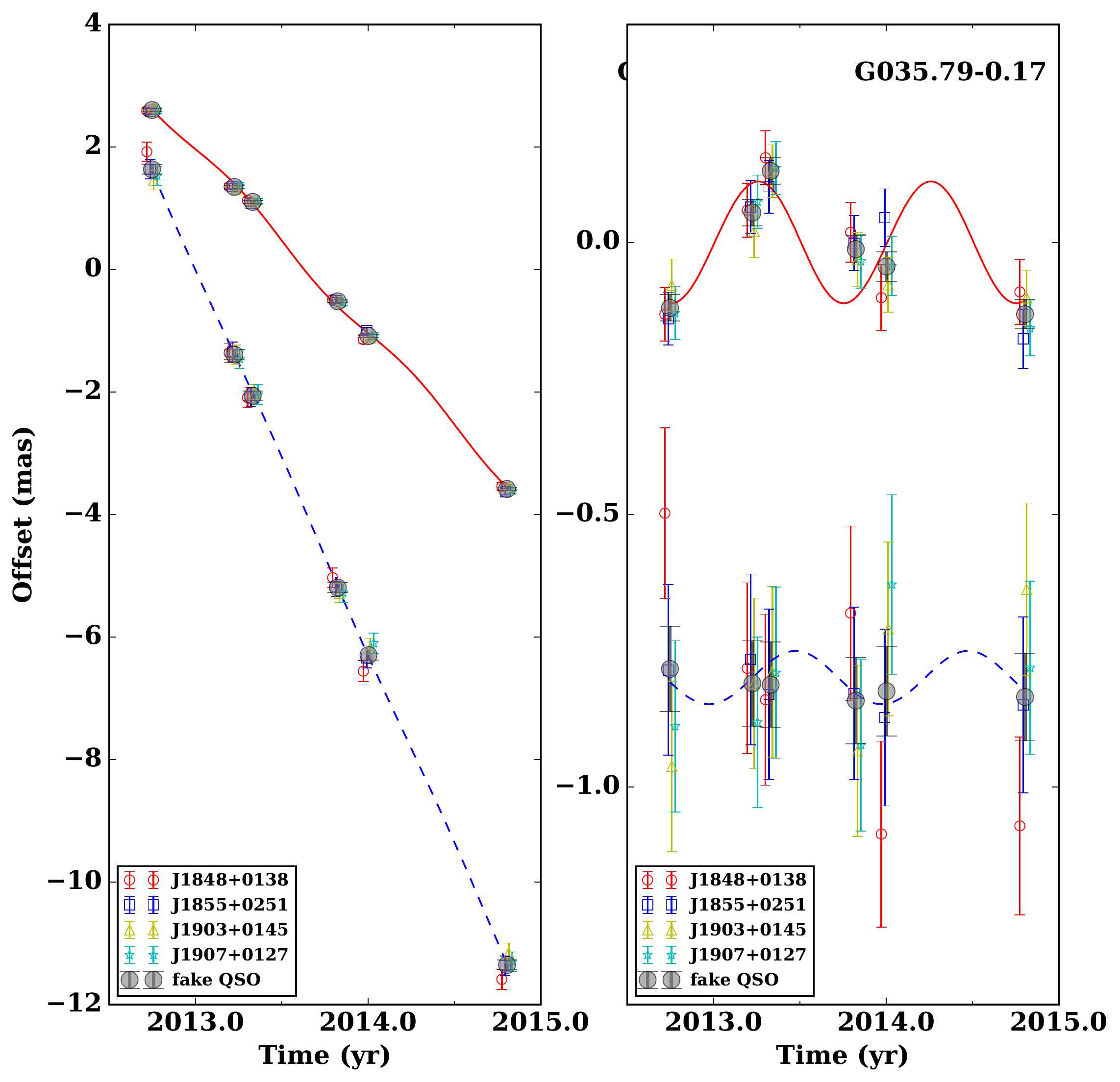}\\
\includegraphics[height=6.5cm]{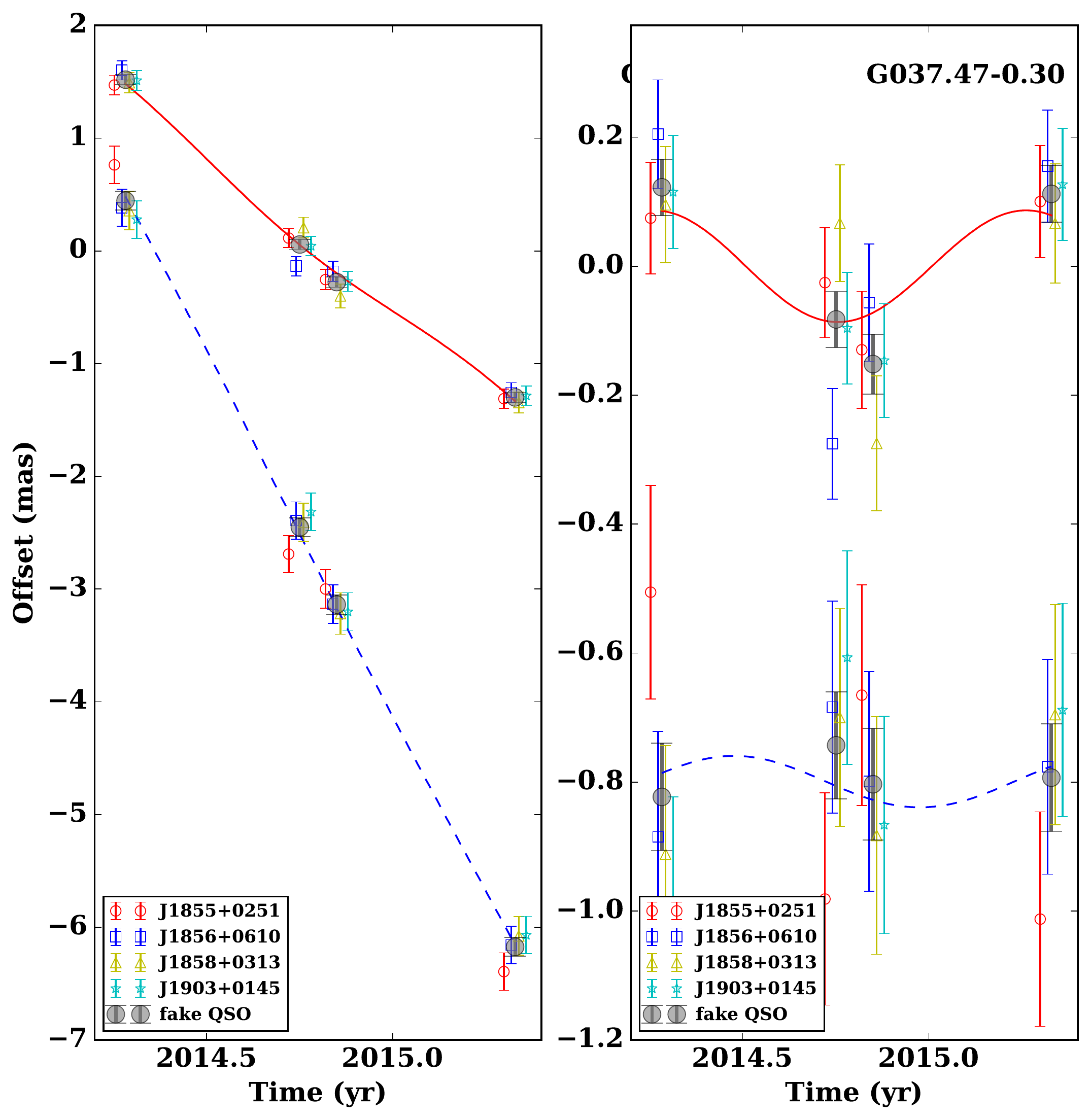}
\includegraphics[height=6.5cm]{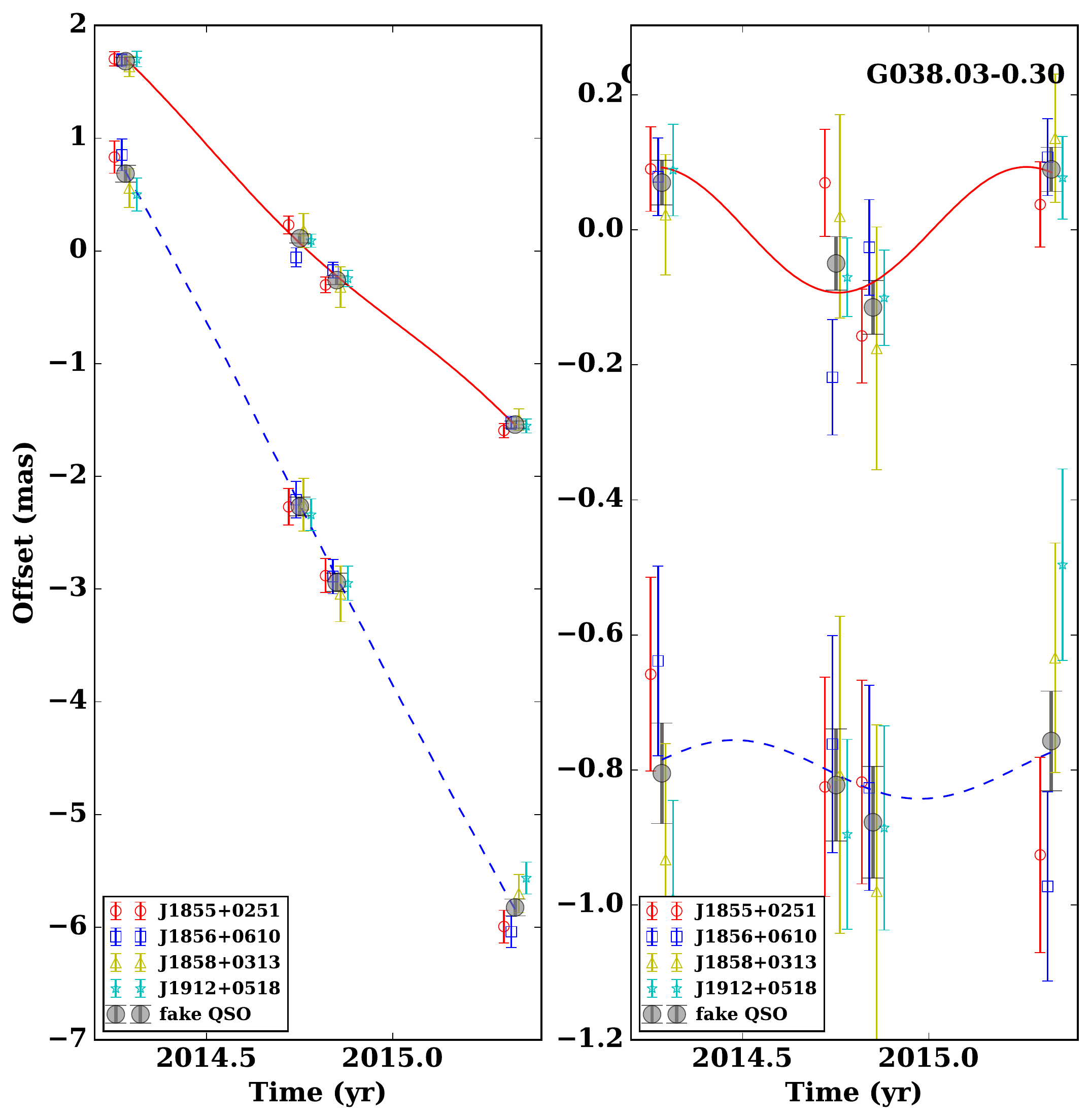}\\
\includegraphics[height=6.5cm]{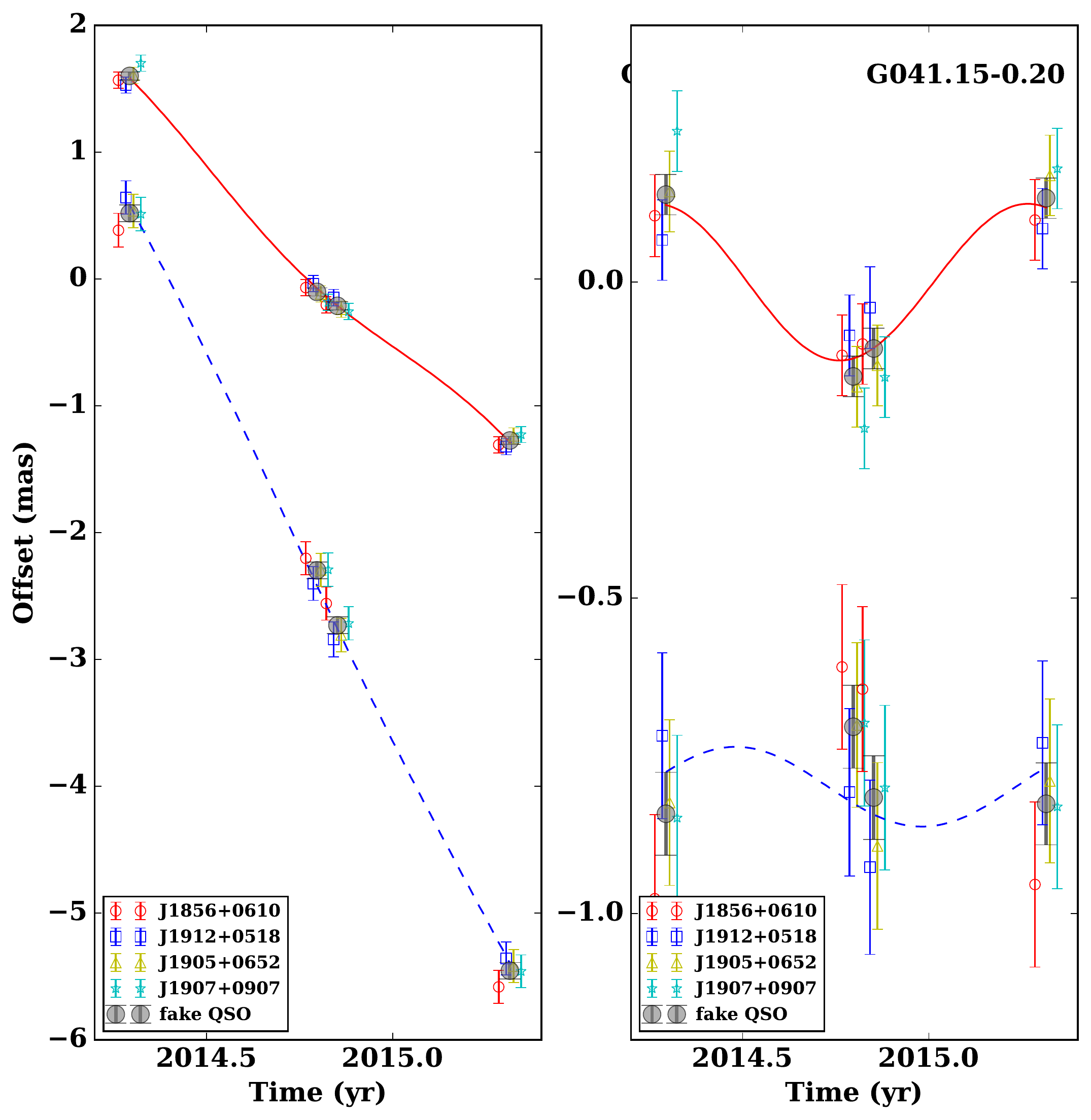}
\includegraphics[height=6.5cm]{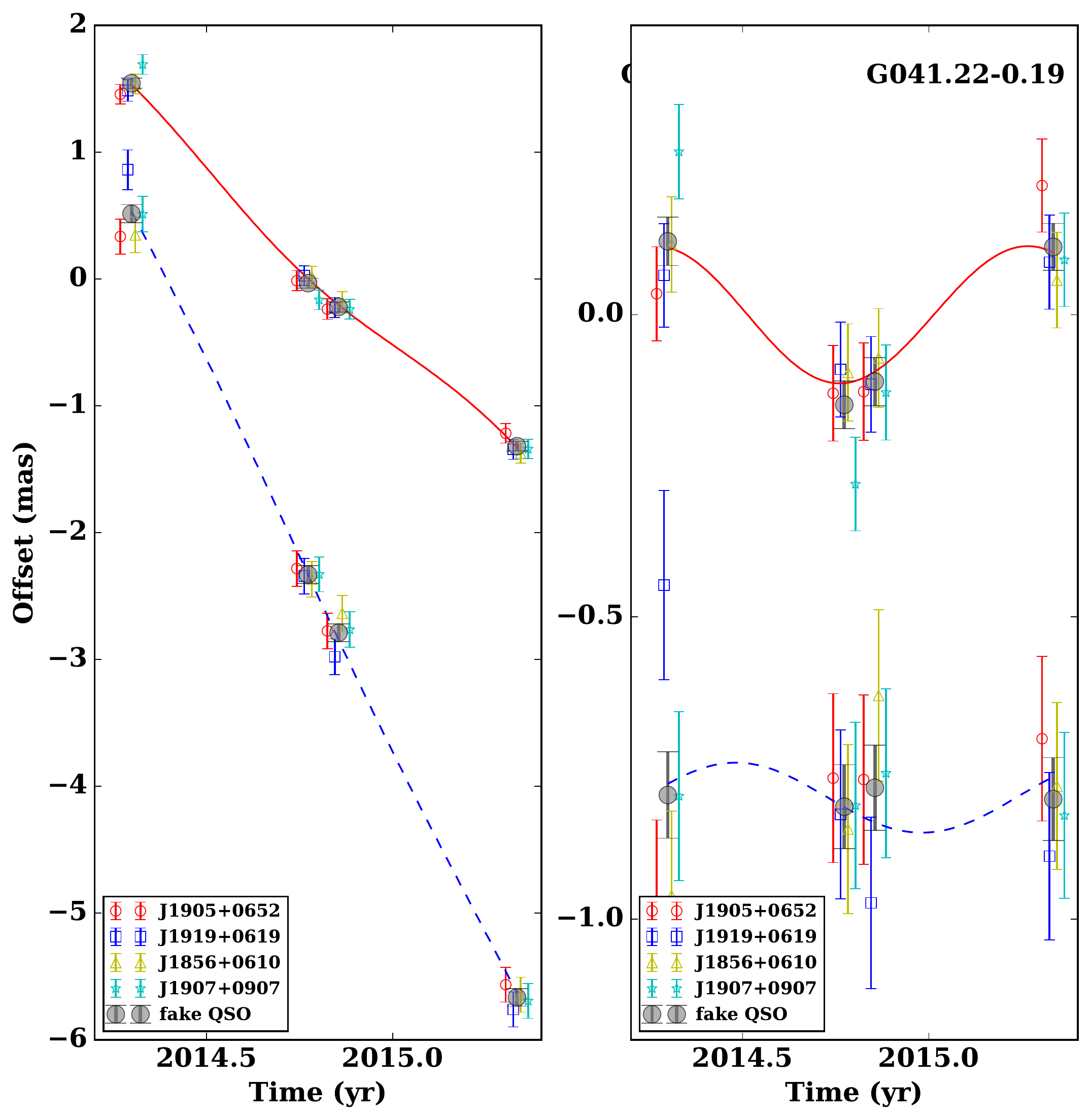}\\
\caption{Parallax and proper motion data and fits.  There are two panels for each
source, with the left panel plotting position vs time and the right panel
has the fitted proper motion removed to better show the parallax signatures.
For methanol masers, the grey circles indicate data for ``artificial quasar''
at the position of the maser. \label{fig-1}}
\end{figure}

\begin{figure}[ht]
\centering
\includegraphics[height=6.5cm]{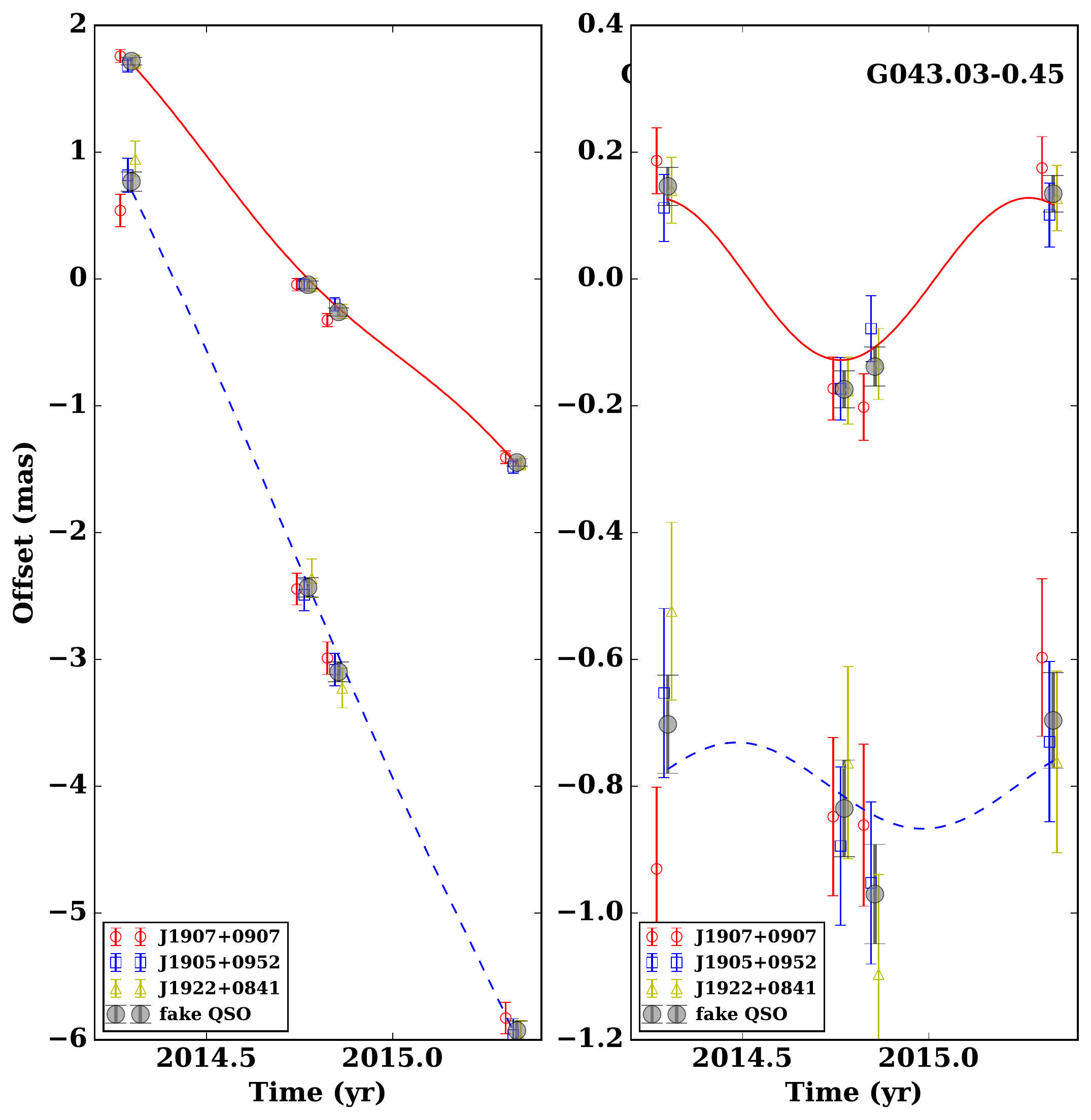}
\includegraphics[height=6.5cm]{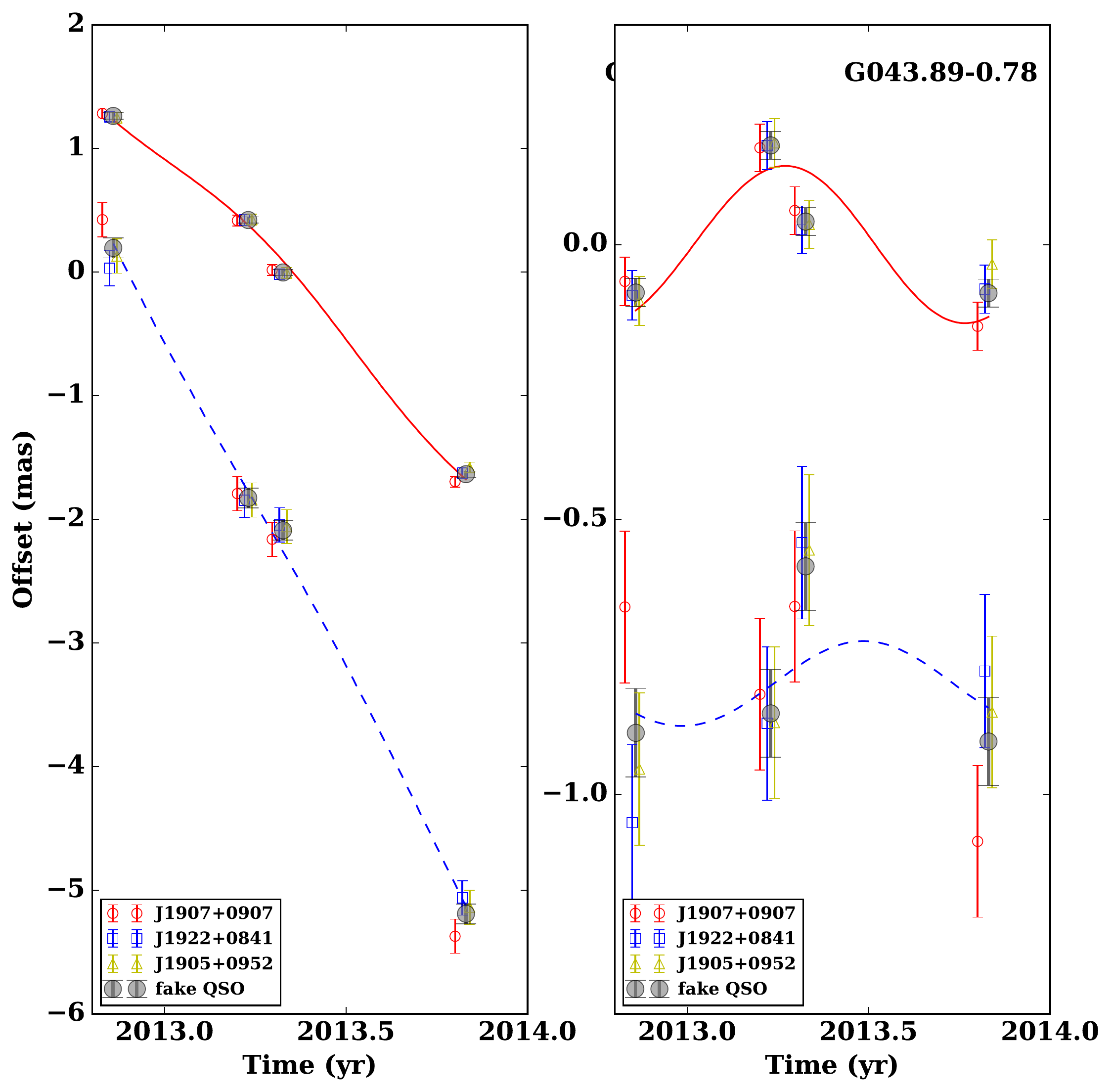}\\
\includegraphics[height=6.5cm]{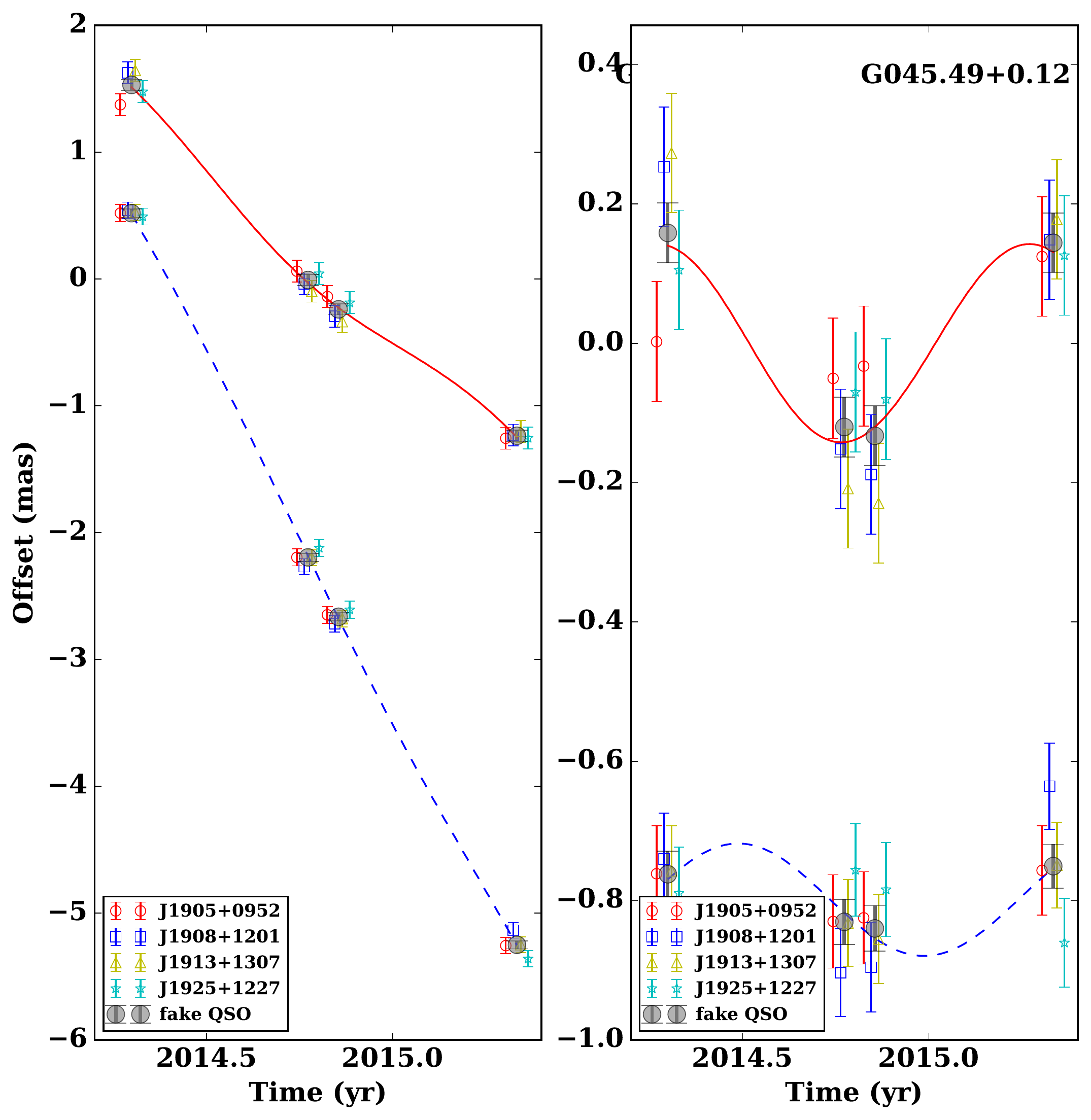}
\includegraphics[height=6.5cm]{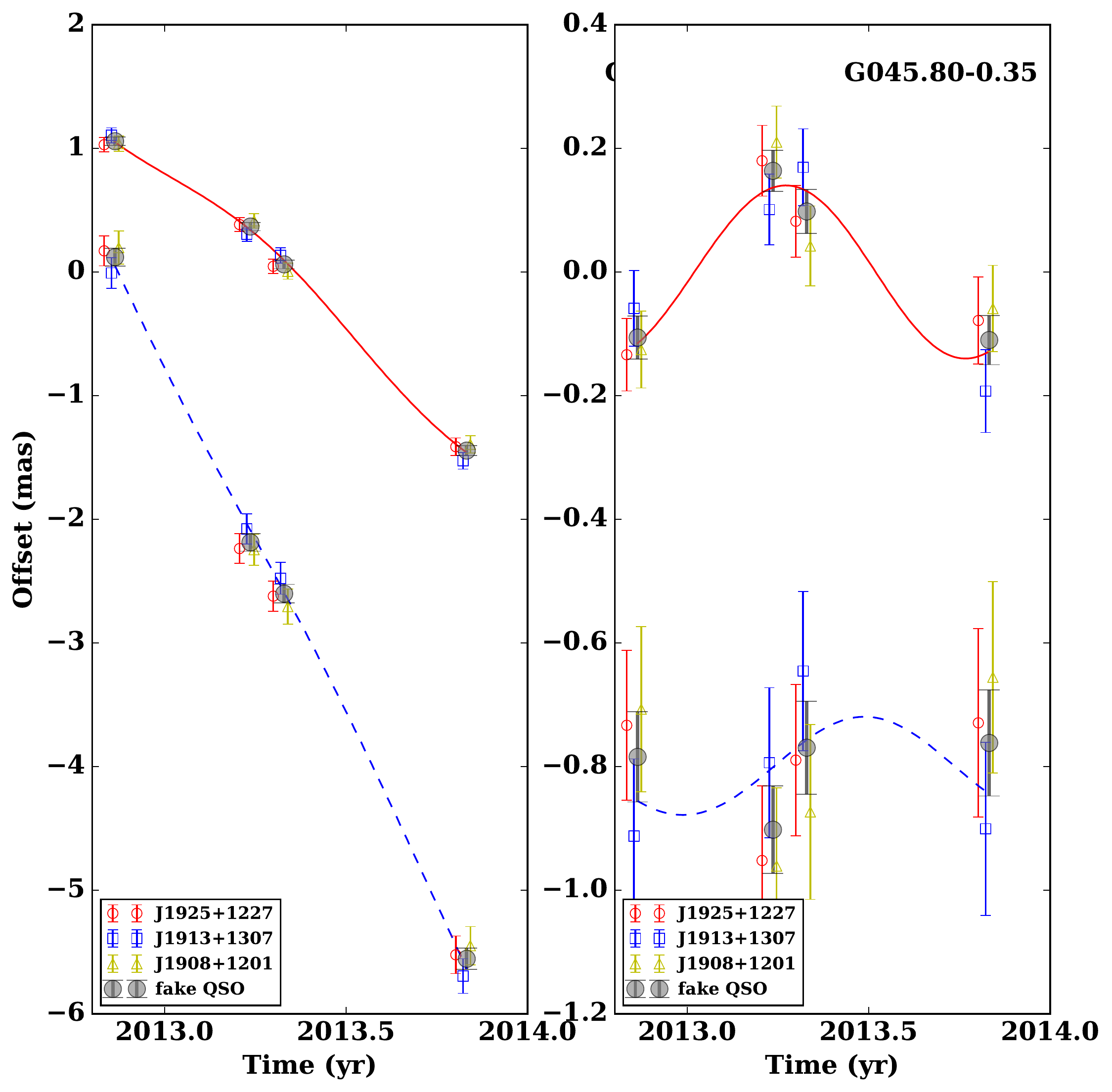}\\
\includegraphics[height=6.5cm]{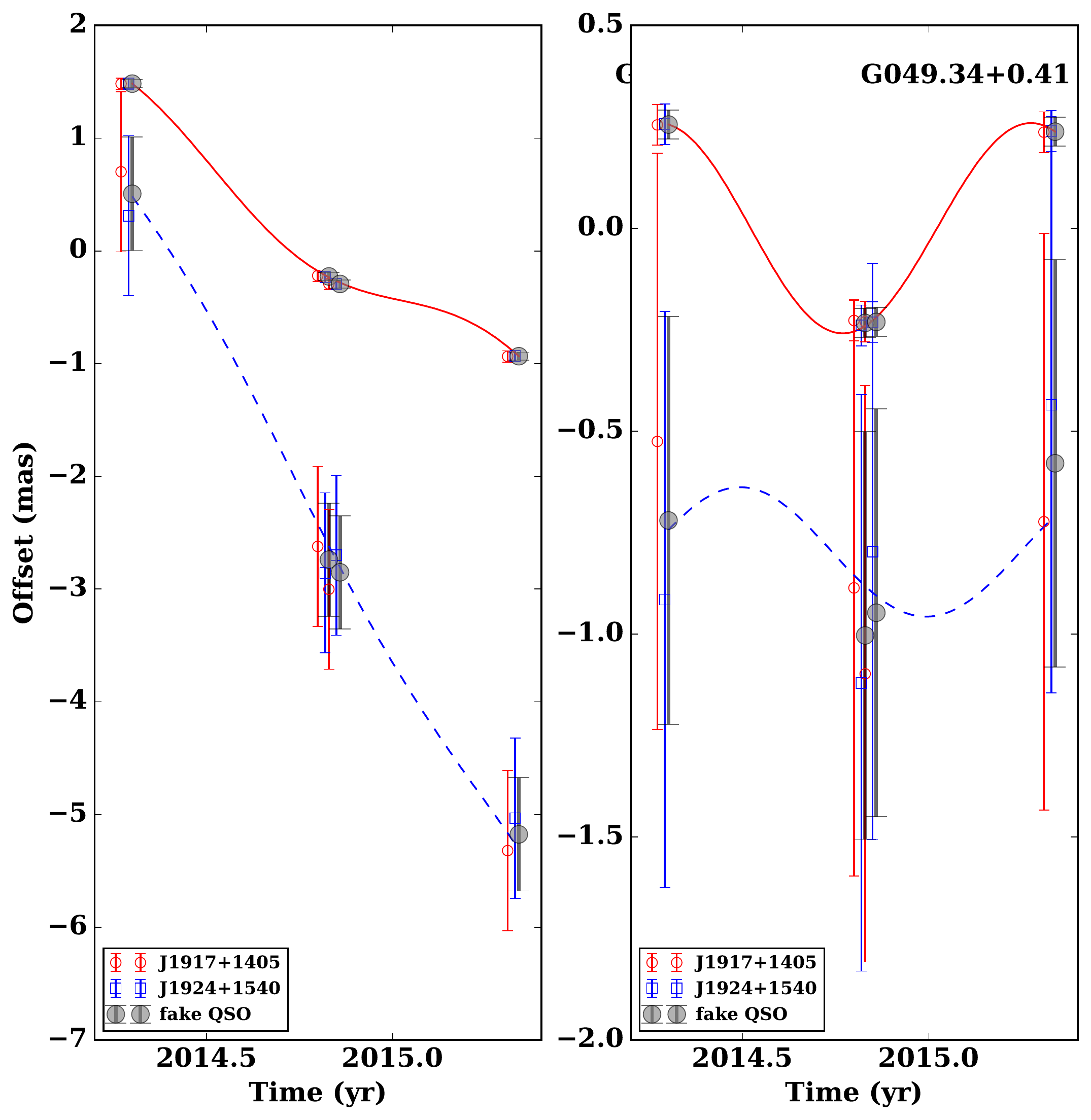}
\includegraphics[height=6.5cm]{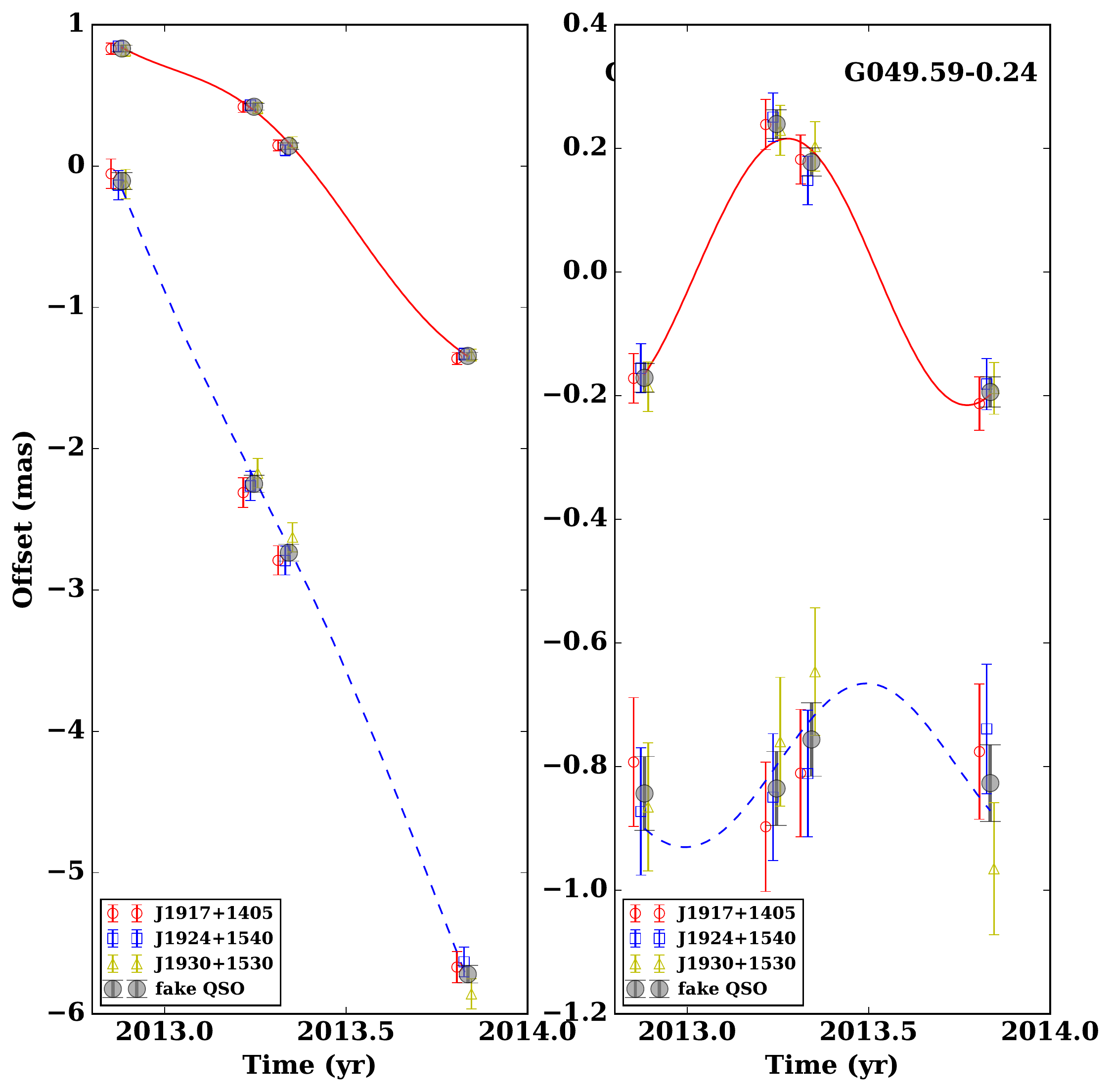}\\
Figure 1 continued
\end{figure}

\begin{figure}[ht]
\centering
\includegraphics[height=6.5cm]{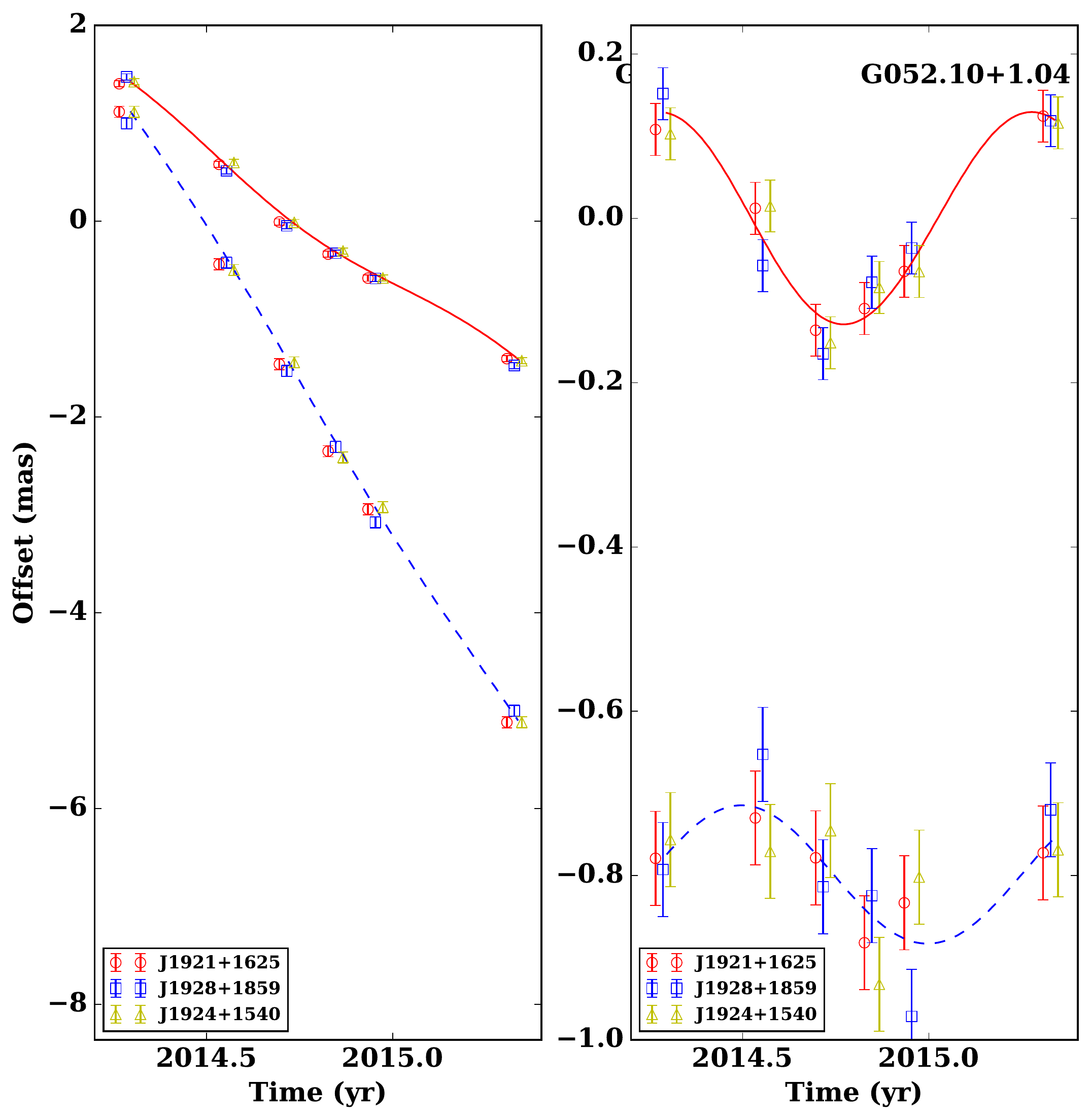}\\
Figure 1 continued
\end{figure}

\clearpage
\begin{deluxetable}{ccccrrrcc}
\tabletypesize{\scriptsize}
\tablecaption{Parallaxes and Proper Motions of HMSFRs in the Sgr Arm Segment
              \label{tbl-1}}
\tablewidth{0pt}
\renewcommand{\arraystretch}{0.8}
\tablehead{
\colhead{Source}  & \colhead{Maser} & \colhead{Parallax} & \colhead{Distance} &
\colhead{$\mu_{x}$}  & \colhead{$\mu_{y}$} & \colhead{V$_{LSR}$} & Ref.\\
name &  & (mas) & (kpc) & (mas yr$^{-1}$) & (mas yr$^{-1}$) &(km s$^{-1}$) & 
}
\startdata
G351.44$+$00.65 &12 GHz      & 0.752 $\pm$ 0.069& $\phantom{1}$1.33$^{+0.13}_{-0.11}$ &   0.31 $\pm$ 0.58 &$-$2.17  $\pm$ 0.90 &$-$8 $\pm$ 3 & 2,11 \\
G011.49$-$01.48 &12 GHz      & 0.800 $\pm$ 0.033& $\phantom{1}$1.25$^{+0.05}_{-0.05}$ &   1.42 $\pm$ 0.52 &$-$0.60  $\pm$ 0.65 &11   $\pm$ 3 & 2 \\
G014.33$-$00.64 &22 GHz      & 0.893 $\pm$ 0.101& $\phantom{1}$1.12$^{+0.14}_{-0.11}$ &   0.95 $\pm$ 1.50 &$-$2.40  $\pm$ 1.30 &22   $\pm$ 5 & 4  \\
G014.63$-$00.57 &22 GHz      & 0.546 $\pm$ 0.022& $\phantom{1}$1.83$^{+0.08}_{-0.07}$ &   0.22 $\pm$ 1.20 &$-$2.07  $\pm$ 1.20 &19   $\pm$ 5 & 2 \\
G015.03$-$00.67 &12 GHz      & 0.499 $\pm$ 0.026& $\phantom{1}$2.00$^{+0.11}_{-0.10}$ &   0.68 $\pm$ 0.53 &$-$1.42  $\pm$ 0.54 &22   $\pm$ 3 & 5,12  \\
G034.39$+$00.22 &22 GHz      & 0.643 $\pm$ 0.049& $\phantom{1}$1.56$^{+0.13}_{-0.11}$ &$-$0.90 $\pm$ 1.00 &$-$2.75  $\pm$ 2.00 &57   $\pm$ 5 & 6  \\
G035.02$+$00.34 &22 GHz      & 0.430 $\pm$ 0.040& $\phantom{1}$2.33$^{+0.24}_{-0.20}$ &$-$0.92 $\pm$ 0.90 &$-$3.61  $\pm$ 0.90 &52   $\pm$ 5 & 2 \\
G035.19$-$00.74 &12 GHz      & 0.456 $\pm$ 0.045& $\phantom{1}$2.19$^{+0.24}_{-0.20}$ &$-$0.18 $\pm$ 0.50 &$-$3.63  $\pm$ 0.50 &30   $\pm$ 7 & 7  \\
G035.20$-$01.73 &12 GHz      & 0.412 $\pm$ 0.014& $\phantom{1}$2.43$^{+0.09}_{-0.08}$ &$-$0.68 $\pm$ 0.44 &$-$3.60  $\pm$ 0.44 &43   $\pm$ 5 & 7,13  \\
G037.42$+$01.51 &12\&22~GHz  & 0.532 $\pm$ 0.021& $\phantom{1}$1.88$^{+0.08}_{-0.07}$ &$-$0.45 $\pm$ 0.35 &$-$3.69  $\pm$ 0.39 &41   $\pm$ 3 & 2 \\
G049.48$-$00.36 & 12 GHz     & 0.195 $\pm$ 0.071& $\phantom{1}$5.13$^{+2.94}_{-1.37}$  &$-$2.49 $\pm$ 0.14 &$-$5.51  $\pm$ 0.16 &56   $\pm$ 3 & 8  \\
G049.48$-$00.38 & 22 GHz     & 0.185 $\pm$ 0.010& $\phantom{1}$5.41$^{+0.31}_{-0.28}$&$-$2.64 $\pm$ 0.20 &$-$5.11  $\pm$ 0.20 &58   $\pm$ 4 & 9  \\
\\
\hline
\\
G032.74$-$00.07 & 22 GHz     & 0.126 $\pm$ 0.016& $\phantom{1}$7.94 $^{+1.15}_{-0.89}$ & -3.15 $\pm$ 0.27 &$-$6.10$\pm$ 0.29 &   37 $\pm$10& 1 \\
G035.79$-$00.17 &6.7~GHz     & 0.113 $\pm$ 0.013& $\phantom{1}$8.85 $^{+1.15}_{-0.91}$ & -2.96 $\pm$ 0.12 &$-$6.23$\pm$ 0.14 &   61 $\pm$ 5& 1 \\
G037.47$-$00.10 &6.7~GHz     & 0.088 $\pm$ 0.030&             11.36 $^{+5.88}_{-2.89}$ & -2.63 $\pm$ 0.07 &$-$6.19$\pm$ 0.15 &   58 $\pm$ 3& 1 \\
G038.03$-$00.30 &6.7~GHz     & 0.095 $\pm$ 0.022&             10.53 $^{+3.17}_{-1.98}$ & -3.01 $\pm$ 0.06 &$-$6.20$\pm$ 0.11 &   60 $\pm$ 3& 1 \\
G041.15$-$00.20 &6.7~GHz     & 0.125 $\pm$ 0.018& $\phantom{1}$8.00 $^{+1.35}_{-1.01}$ & -2.79 $\pm$ 0.14 &$-$5.85$\pm$ 0.16 &   60 $\pm$ 3& 1 \\
G041.22$-$00.19 &6.7~GHz     & 0.113 $\pm$ 0.022& $\phantom{1}$8.85 $^{+2.14}_{-1.44}$ & -2.82 $\pm$ 0.13 &$-$5.89$\pm$ 0.16 &   59 $\pm$ 5& 1 \\
G043.03$-$00.45 &6.7~GHz     & 0.130 $\pm$ 0.019& $\phantom{1}$7.69 $^{+1.32}_{-0.98}$ & -3.03 $\pm$ 0.15 &$-$6.56$\pm$ 0.20 &   56 $\pm$ 3& 1 \\
G043.79$-$00.12 &22  GHz     & 0.166 $\pm$ 0.010& $\phantom{1}$6.02 $^{+0.39}_{-0.34}$ & -3.02 $\pm$ 0.36 &$-$6.20$\pm$ 0.36 &   44 $\pm$10& 2 \\
G043.89$-$00.78 &6.7\&22~GHz & 0.134 $\pm$ 0.013& $\phantom{1}$7.41 $^{+0.79}_{-0.65}$ & -3.11 $\pm$ 0.15 &$-$5.35$\pm$ 0.21 &   50 $\pm$ 3& 1,2 \\
G045.07$+$00.13 &22  GHz     & 0.129 $\pm$ 0.007& $\phantom{1}$7.75 $^{+0.45}_{-0.40}$ & -3.21 $\pm$ 0.26 &$-$6.11$\pm$ 0.26 &   59 $\pm$ 5& 2 \\
G045.45$+$00.06 &22  GHz     & 0.119 $\pm$ 0.017& $\phantom{1}$8.40 $^{+1.40}_{-1.05}$ & -2.34 $\pm$ 0.38 &$-$6.00$\pm$ 0.54 &   55 $\pm$ 7& 2 \\
G045.49$+$00.12 &6.7~GHz     & 0.144 $\pm$ 0.024& $\phantom{1}$6.94 $^{+1.39}_{-0.99}$ & -2.62 $\pm$ 0.17 &$-$5.61$\pm$ 0.16 &   57 $\pm$ 3& 1 \\
G045.80$-$00.35 &6.7~GHz     & 0.137 $\pm$ 0.023& $\phantom{1}$7.30 $^{+1.47}_{-1.05}$ & -2.52 $\pm$ 0.17 &$-$6.08$\pm$ 0.27 &   64 $\pm$ 5& 1 \\
G048.99$-$00.29 &22  GHz     & 0.178 $\pm$ 0.017& $\phantom{1}$5.62 $^{+0.59}_{-0.49}$ & -2.20 $\pm$ 0.48 &$-$5.84$\pm$ 0.63 &   67 $\pm$10& 3 \\
G049.19$-$00.33 &22  GHz     & 0.197 $\pm$ 0.008& $\phantom{1}$5.08$^{+0.22}_{-0.20}$  & -3.08 $\pm$ 0.40 &$-$5.50$\pm$ 0.40 &   67 $\pm$ 5& 2, 3\\
G049.34$+$00.41 &6.7~GHz     & 0.241 $\pm$ 0.031& $\phantom{1}$4.15$^{+0.61}_{-0.47}$  & -2.36 $\pm$ 0.26 &$-$5.59$\pm$ 0.33 &   68 $\pm$ 5& 1 \\
G049.59$-$00.24 &6.7~GHz     & 0.218 $\pm$ 0.009& $\phantom{1}$4.59 $^{+0.20}_{-0.18}$ & -2.25 $\pm$ 0.24 &$-$6.12$\pm$ 0.25 &   63 $\pm$ 5& 1 \\
G052.10$+$01.04 &22  GHz     & 0.165 $\pm$ 0.013& $\phantom{1}$6.06 $^{+0.52}_{-0.44}$ & -2.77 $\pm$ 1.40 &$-$5.85$\pm$ 1.40 &   42 $\pm$40& 1,10 \\
\enddata
\tablecomments{Column~1 gives the source name.
Column~2 gives the maser transition: 6.7 and 12.2~GHz are CH$_3$OH and 22 GHz are H$_2$O masers.
Columns~3 and 4 give trigonometric parallaxes and distances.
Columns~5, 6 and 7 give the proper motion components in the eastward
($\mu_x$=$\mu_\alpha$~cos$\delta$) and northward directions ($\mu_y$ =
$\mu_\delta$) and the estimated LSR velocity of the central exciting star.
References are given in Column 8. The dividing line within the Table separates Sgr Far
sources (below line) from Sgr Near sources (above line).} 
 \tablerefs{
 (1) this paper;
 (2) \citeauthor{2014AA...566A..17W} \citeyear{2014AA...566A..17W}. 
 (3) \citeauthor{2015PASJ...67...65N} \citeyear{2015PASJ...67...65N}. 
 (4) \citeauthor{Sato2010} \citeyear{Sato2010};
 (5) \citeauthor{Xu2011} \citeyear{Xu2011};
 (6) \citeauthor{Kurayama2011} \citeyear{Kurayama2011};
 (7) \citeauthor{Zhang2009} \citeyear{Zhang2009};
 (8) \citeauthor{Xu2009} \citeyear{Xu2009};
 (9) \citeauthor{Sato2010a} \citeyear{Sato2010a};
 (10) \citeauthor{Oh2010} \citeyear{Oh2010}
 (11) \citeauthor{2014PASJ...66..104C} \citeyear{2014PASJ...66..104C}
 (12) \citeauthor{2016MNRAS.460.1839C} \citeyear{2016MNRAS.460.1839C}
 (13) \citeauthor{2018Rygl} \citeyear{2018Rygl}
 }
\end{deluxetable}
\clearpage
\begin{deluxetable}{ccrrr}
\tabletypesize{\scriptsize}
\tablecaption{Peculiar Motions \label{tbl-2}}
\tablewidth{0pt}
\renewcommand{\arraystretch}{0.8}
\tablehead{
\colhead{Source} &\colhead{Arm}  & \colhead{U$_s$} & \colhead{V$_s$} & \colhead{W$_s$} \\
Name & Section &(km s$^{-1}$)& (km s$^{-1}$) & (km s$^{-1}$)
}
\startdata
G351.44$+$00.65  &Near&  $\phantom{1}+0.3\pm\phantom{1}3.1$ & $\phantom{1}+6.4\pm\phantom{1}5.2$  & $\phantom{1}-0.7\pm\phantom{1}4.6$ \\
G011.49$-$01.48  &Near&  $\phantom{1}-0.8\pm\phantom{1}3.0$ &           $+14.5\pm\phantom{1}3.6$  & $\phantom{1}-0.1\pm\phantom{1}3.3$ \\
G014.33$-$00.64  &Near&            $+11.1\pm\phantom{1}5.4$ & $\phantom{1}+8.0\pm\phantom{1}7.2$  & $\phantom{1}-1.7\pm\phantom{1}8.0$ \\
G014.63$-$00.57  &Near&  $\phantom{1}+2.1\pm\phantom{1}5.9$ & $\phantom{1}+1.9\pm10.0$            & $\phantom{1}-1.3\pm10.4$           \\
G015.03$-$00.67  &Near&  $\phantom{1}+0.3\pm\phantom{1}3.5$ & $\phantom{1}+8.9\pm\phantom{1}4.9$  & $\phantom{1}-3.2\pm\phantom{1}5.1$ \\
G034.39$+$00.22  &Near&            $+22.6\pm10.2$           &           $+23.9\pm11.0$            & $\phantom{1}+5.7\pm\phantom{1}9.6$ \\
G035.02$+$00.34  &Near&            $+11.3\pm\phantom{1}8.8$ & $\phantom{1}+9.0\pm\phantom{1}8.1$  & $\phantom{1-}0.0\pm10.0$           \\
G035.19$-$00.74  &Near&  $\phantom{1}-4.6\pm\phantom{1}7.0$ & $\phantom{1}-3.2\pm\phantom{1}6.7$  & $\phantom{1}-6.8\pm\phantom{1}5.5$ \\
G035.20$-$01.73  &Near&  $\phantom{1}+2.4\pm\phantom{1}5.2$ & $\phantom{1}+2.4\pm\phantom{1}5.1$  & $\phantom{1}-3.9\pm\phantom{1}5.1$ \\
G037.42$+$01.51  &Near&  $\phantom{1}+7.1\pm\phantom{1}3.4$ & $\phantom{1}+8.3\pm\phantom{1}3.4$  & $\phantom{1}-1.9\pm\phantom{1}3.2$ \\
G049.48$-$00.36  &Near&  $\phantom{1}-0.7\pm18.6$           & $\phantom{1}-1.8\pm41.2$            & $\phantom{1}-1.3\pm\phantom{1}7.6$ \\
G049.48$-$00.38  &Near&            $-10.5\pm\phantom{1}6.1$ & $\phantom{+1}0.0\pm\phantom{1}4.0$  & $\phantom{1}+6.3\pm\phantom{1}5.1$ \\
\\
\hline
\\
G032.74$-$00.07  &Far &            $+14.2\pm12.0$           &           $-59.0\pm\phantom{1}16.2$ & $ \phantom{1}+9.7\pm10.7$           \\
G035.79$-$00.17  &Far &  $\phantom{1}+1.2\pm\phantom{1}9.6$ &           $-17.2\pm\phantom{1}15.9$ & $ \phantom{1}-0.1\pm\phantom{1}5.4$ \\
G037.47$-$00.10  &Far &            $-19.5\pm15.1$           &           $+21.0\pm113.9$           &            $-17.6\pm12.1$           \\
G038.03$-$00.30  &Far &  $\phantom{1}-9.4\pm12.4$           &           $+15.4\pm\phantom{1}72.7$ & $ \phantom{1}-0.6\pm\phantom{1}4.7$ \\
G041.15$-$00.20  &Far &  $\phantom{1}-8.3\pm10.6$           &           $-12.9\pm\phantom{1}14.6$ & $ \phantom{1}+0.6\pm\phantom{1}5.8$ \\
G041.22$-$00.19  &Far &            $-12.2\pm12.6$           & $\phantom{1}-5.4\pm\phantom{1}36.7$ & $ \phantom{1}-0.1\pm\phantom{1}6.4$ \\
G043.03$-$00.45  &Far &            $+20.6\pm\phantom{1}9.0$ & $\phantom{1}-5.1\pm\phantom{1}19.4$ & $ \phantom{1}-4.1\pm\phantom{1}6.4$ \\
G043.79$-$00.12  &Far &            $+19.2\pm10.5$           &           $-28.9\pm\phantom{1}10.0$ & $ \phantom{1}+3.4\pm10.4$           \\
G043.89$-$00.78  &Far &            $-11.6\pm\phantom{1}8.7$ &           $-17.3\pm\phantom{11}5.9$ &            $+17.7\pm\phantom{1}5.9$ \\
G045.07$+$00.13  &Far &  $\phantom{1}+7.5\pm\phantom{1}9.7$ & $\phantom{1}+2.6\pm\phantom{11}7.3$ & $ \phantom{1}+9.4\pm\phantom{1}9.6$ \\
G045.45$+$00.06  &Far &            $-13.3\pm20.0$           & $\phantom{1}+0.1\pm\phantom{1}20.4$ &            $-19.6\pm17.8$           \\
G045.49$+$00.12  &Far &            $-10.4\pm11.3$           & $\phantom{1}-9.6\pm\phantom{1}11.2$ & $ \phantom{1}-0.4\pm\phantom{1}6.1$ \\
G045.80$-$00.35  &Far &  $\phantom{1}-2.9\pm13.1$           & $\phantom{1}+2.6\pm\phantom{1}17.1$ &            $-12.3\pm\phantom{1}8.2$ \\
G048.99$-$00.29  &Far &  $\phantom{1}-1.0\pm16.9$           & $\phantom{1}+7.7\pm\phantom{1}10.2$ &            $-13.0\pm14.4$           \\
G049.19$-$00.33  &Far &  $\phantom{1}+7.1\pm\phantom{1}9.9$ & $\phantom{1}+8.3\pm\phantom{11}5.0$ & $           +11.5\pm\phantom{1}9.7$ \\
G049.34$+$00.41  &Far &            $+10.7\pm\phantom{1}8.8$ &           $+12.3\pm\phantom{11}5.4$ & $ \phantom{1}-1.4\pm\phantom{1}5.8$ \\
G049.59$-$00.24  &Far &            $+14.1\pm\phantom{1}5.6$ & $\phantom{1}+5.3\pm\phantom{11}5.0$ &            $-11.1\pm\phantom{1}5.4$ \\ 
G052.10$+$1.04  &Far &  $\phantom{1}+3.8\pm\phantom{1}7.9$ & $\phantom{1}-6.5\pm\phantom{1}39.4$ & $ \phantom{1}+0.1\pm\phantom{1}4.2$ \\
\\
\hline                                   
\\
\bf{Average}&\bf{Near} & \bf{2.0 $\pm$ 1.4}& \bf{7.1$ \pm$ 1.5} &\bf{$-$1.1 $\pm$ 1.5}\\
\bf{Average}&\bf{Far}  & \bf{4.3 $\pm$ 2.4}& \bf{$-$0.6$ \pm$ 2.2} &\bf{$-$0.2 $\pm$ 1.6}\\
\bf{Average}&\bf{All}  & \bf{2.6 $\pm$ 1.1}& \bf{4.5$ \pm$ 1.3} &\bf{$-$0.7 $\pm$ 1.1}\\
\enddata          
\tablecomments{$U_s, V_s$, $W_s$ give the estimated peculiar (non-circular)
velocity components of the central star that excites the masers, assuming R$_0$
= 8.31, $\Theta_0$ = 241 km s$^{-1}$, and Solar Motion components (U$\odot$,
V$\odot$, W$\odot$) = (10.5, 14.4, 8.9)~km~s$^{-1}$.}
\end{deluxetable}

\appendix
\setcounter{figure}{0} \renewcommand{\thefigure}{A.\arabic{figure}}
\setcounter{table}{0} \renewcommand{\thetable}{A.\arabic{table}}

\begin{figure}[ht]
\centering
\includegraphics[width=15cm]{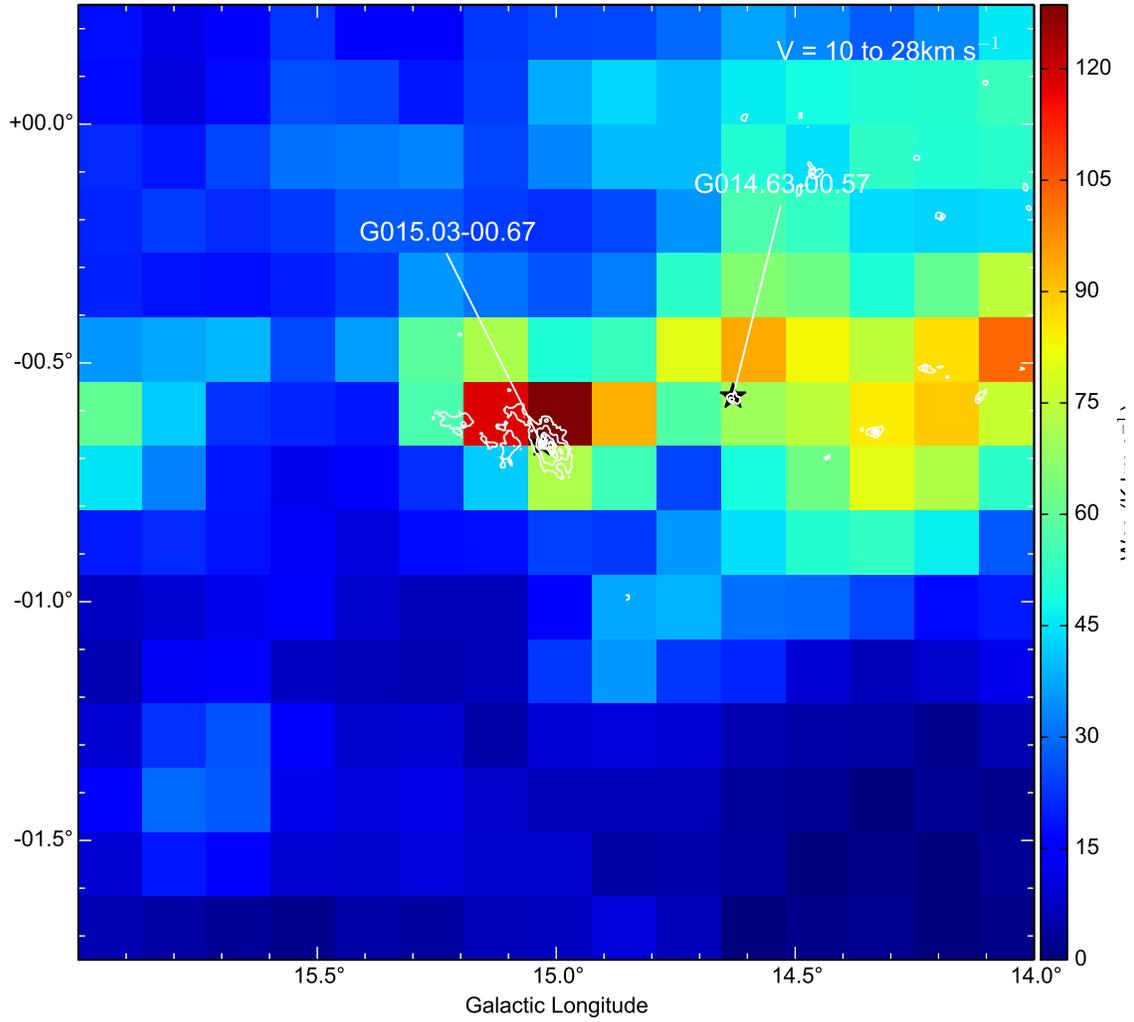}\\
\caption{False color image the of CO (J = 1$-$0) emission
integrated over 10 - 28 km s$^{-1}$ from the CfA 1.2-m Survey 
\citep{Dame2001}. Over plotted are
contours of 870~$\mu$m dust emission from the
ATLASGAL survey \citep{Schuller2009}, with contour levels of 1, 3, 7,11, 15, 19
Jy~beam$^{-1}$. The positions of the masers are indicated with black stars.
 \label{fig-G15.03_CO}\label{fig-A1}}
\end{figure}

\begin{figure}[ht]
\centering
\includegraphics[width=15cm]{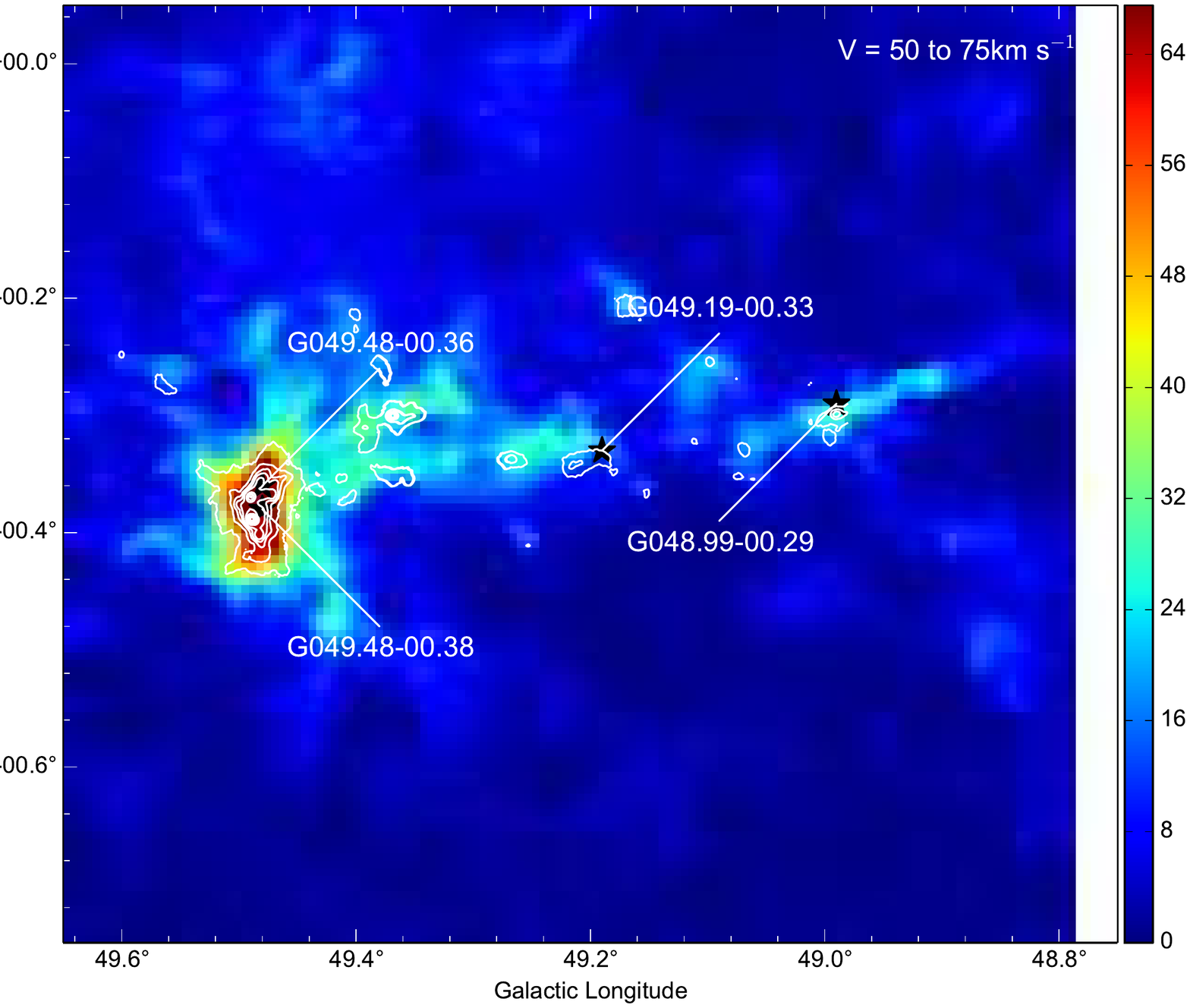}\\
\caption{False color image of $^{13}$CO (J = 1$-$0) emission from the GRS
$^{13}$CO Survey Archive \citep[][]{2006ApJS..163..145J} integrated over 50 $<$
V$_{LSR}$ $<$ 75 km s$^{-1}$. Over plotted are contours of 870~$\mu$m dust
emission from the ATLASGAL survey \citep{Schuller2009}, with contour levels of
0.5, 1, 3, 5 7, 10, 20, ..., 70 Jy beam$^{-1}$. The position of masers are
indicated with black stars.  \label{fig-A2}} \end{figure}

\begin{figure}[ht]
\centering
\includegraphics[width=15cm]{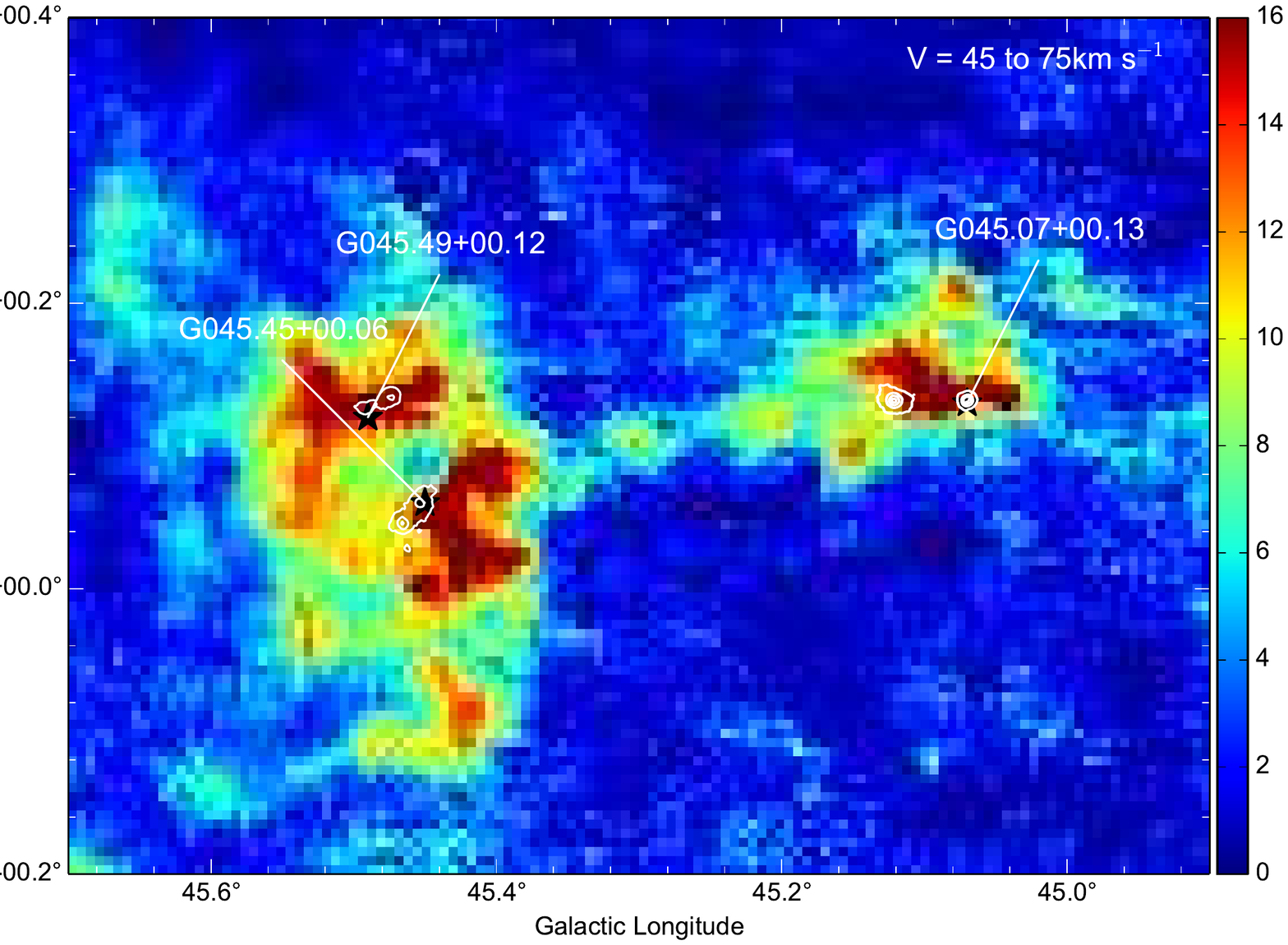}\\
\caption{False color image of $^{13}$CO (J = 1$-$0)
emission from the GRS $^{13}$CO Survey Archive \citep[][]{2006ApJS..163..145J} integrated over 45 $<$ V$_{LSR}$
$<$ 75 km s$^{-1}$. Over plotted are contours of 870~$\mu$m dust emissions from the
ATLASGAL survey \citep{Schuller2009}, with contour levels of  0.2, 0.5, 1, 3, 5
and 7 Jy~beam$^{-1}$. The position of masers are indicated with black
stars.\label{fig-A3}}
\end{figure}

\begin{deluxetable}{llrcccccc}
\tabletypesize{\scriptsize}
\tablecaption{Observations \label{tbl-A1}}
\tablewidth{0pt}
\tablehead{
\colhead{Source} & \colhead{Project}  & \colhead{Maser} & \colhead{Epoch 1} & \colhead{Epoch 2} &
\colhead{Epoch 3} & \colhead{Epoch 4} & \colhead{Epoch 5} & \colhead{Epoch 6} } \startdata
G032.74$-$00.07 & BR198N &  H$_2$O    & 2014Apr11 & 2014Jul21 & 2014Sep19 & 2014Oct29 & 2014Nov30 & 2015Apr14 \\
G035.79$-$00.17 & BR149N &  CH$_3$OH  & 2012Sep29 & 2013Mar22 & 2013Apr30 & 2013Oct27 & 2013Dec31 & 2014Oct20 \\
G037.47$-$00.10 & BR198P &  CH$_3$OH  & 2014Apr13 & 2014Oct01 & 2014Nov06 & 2015Apr30 \\
G038.03$-$00.30 & BR198P &  CH$_3$OH  & 2014Apr13 & 2014Oct01 & 2014Nov06 & 2015Apr30 \\
G041.15$-$00.20 & BR198Q &  CH$_3$OH  & 2014Apr17 & 2014Oct18 & 2014Nov07 & 2015Apr25 \\
G041.22$-$00.19 & BR198R &  CH$_3$OH  & 2014Apr19 & 2014Oct09 & 2014Nov08 & 2015May02 \\
G043.03$-$00.45 & BR198R &  CH$_3$OH  & 2014Apr19 & 2014Oct09 & 2014Nov08 & 2015May02 \\
G043.89$-$00.78 & BR149O &  CH$_3$OH  & 2012Nov08 & 2013Mar24 & 2013Apr28 & 2013Oct29 \\
G045.49$+$00.12 & BR198R &  CH$_3$OH  & 2014Apr19 & 2014Oct09 & 2014Nov08 & 2015May02 \\
G045.80$-$00.35 & BR149P &  CH$_3$OH  & 2012Nov10 & 2013Mar26 & 2013Apr29 & 2013Oct30 \\
G049.34$+$00.41 & BR198T &  CH$_3$OH  & 2014Apr20 & 2014Oct30 & 2014NoV10 & 2015May04 \\
G052.10$+$01.04 & BR198S &   H$_2$O   & 2014Apr18 & 2014Jul25 & 2014Sep22 & 2014Nov09 & 2014DEC18& 2015MAY03\\
\enddata
\end{deluxetable}

\begin{deluxetable}{lcccrlcr}
\tabletypesize{\footnotesize}
\rotate
\tablecaption{1st epoch Source information \label{tbl-A2}}
\tablewidth{0pt}
\tablehead{
\colhead{Source} & \colhead{R.A.(J2000)} & \colhead{Dec.(J2000)} &
\colhead{R.A.$_{sep}$}  & \colhead{DEC.$_{sep}$} & \colhead{Beam Size} & \colhead{F$_{peak}$}
 & \colhead{V$_{peak}$} \\
       & (h m s) & ($^{\circ}$,$^{\prime}$,$^{\prime\prime}$) & ($^{\circ}$) &
($^{\circ}$) & (mas $\times$ mas at $^\circ$) & (Jy beam$^{-1}$) &  (km s$^{-1}$)
}
\startdata
G032.74$-$00.07 & 18:51:21.8627  & $-$00:12:06.3359 & ...  & ...  & 1.28 $\times$ 0.82 at -23 &  9.18 &$+$33.0 \\
J1857$-$0048    & 18:57:51.35860 & $-$00:48:21.9496 & 0.45 & 0.02 & 2.42 $\times$ 0.90 at -46 & 0.009 &   ...  \\
J1853$-$0010    & 18:53:10.26920 & $-$00:10:50.7400 & 1.62 &-0.60 & 2.54 $\times$ 0.91 at -46 & 0.008 &   ...  \\
J1848$+$0138    & 18:48:21.81035 & $+$01:38:26.6322 &-0.75 & 1.84 & 1.09 $\times$ 0.79 at -16 & 0.036 &   ...  \\
\tableline
G035.79$-$00.17 &18:57:16.89050  & $+$02:27:58.0070 & ...  & ... & 3.17 $\times$ 1.20 at -2 &  4.07 &$+$61.00 \\
1855+0251       &18:55:35.43640  & $+$02:51:19.5629 &-0.42 & 0.39& 3.03 $\times$ 1.25 at  1 &  0.043&   ...  \\
1903+0145       &19:03:53.06320  & $+$01:45:26.3108 & 1.65 &-0.71& 3.04 $\times$ 1.24 at  0 &  0.072&   ...  \\
1848+0138       &18:48:21.81030  & $+$01:38:26.6322 &-2.23 &-0.83& 3.06 $\times$ 1.25 at  0 &  0.025&   ...  \\
1907+0127       &19:07:11.99610  & $+$01:27:08.9644 & 2.48 &-1.01& 3.08 $\times$ 1.28 at  2 &  0.086&   ...  \\
\tableline
G037.47$-$00.10 &19:00:07.14300  & $+$03:59:52.9750 &      &     & 4.52 $\times$ 2.68 at 12 &  2.87 &$+$62.00 \\
J1855+0251      &18:55:35.43640  & $+$02:51:19.5630 &-1.13 &-1.14& 5.66 $\times$ 2.39 at 25 &  0.065&   ...  \\
J1858+0313      &18:58:02.35227  & $+$03:13:16.3116 &-0.52 &-0.78& 4.71 $\times$ 2.85 at 18 &  0.137&   ...  \\
J1856+0610      &18:56:31.83880  & $+$06:10:16.7650 &-0.89 & 2.17& 4.57 $\times$ 2.82 at  0 &  0.147&   ...  \\
J1903+0145      &19:03:53.06327  & $+$01:45:26.3113 & 0.94 &-2.24& 4.29 $\times$ 2.52 at -2 &  0.052&   ...  \\
\tableline
G038.03$-$00.30 &19:01:50.46760  & $+$04:24:18.8999 &      &     & 3.76 $\times$ 2.18 at  4 &  3.10 &$+$56.00 \\
J1855+0251      &18:55:35.43640  & $+$02:51:19.5630 &-1.56 &-1.55& 4.15 $\times$ 2.47 at 16 &  0.060&   ...  \\
J1856+0610      &18:56:31.83880  & $+$06:10:16.7650 &-1.32 & 1.77& 3.64 $\times$ 2.08 at -2 &  0.138&   ...  \\
J1912+0518      &19:12:54.25770  & $+$05:18:00.4220 & 2.76 & 0.89& 3.73 $\times$ 2.24 at -3 &  0.047&   ...  \\
J1858+0313      &18:58:02.35227  & $+$03:13:16.3116 &-0.95 &-1.18& 3.74 $\times$ 1.95 at  5 &  0.092&   ...  \\
\tableline
G041.15$-$00.20 & 19:07:14.37300 & $+$07:13:17.6500 & ...  & ... & 3.11 $\times$ 1.37 at 1 &  3.26 &$+$56.0 \\
J1905+0652      & 19:05:21.21048 & $+$06:52:10.7803 &-0.47 &-0.35& 3.23 $\times$ 1.47 at 0 & 0.092 &   ...  \\
J1907+0907      & 19:07:41.96340 & $+$09:07:12.3970 & 0.11 & 1.90& 3.19 $\times$ 1.46 at 0 & 0.186 &   ...  \\
J1856+0610      & 18:56:31.83880 & $+$06:10:16.7650 &-2.66 &-1.05& 3.21 $\times$ 1.46 at 0 & 0.030 &   ...  \\
J1912+0518      & 19:12:54.25770 & $+$05:18:00.4220 & 1.40 &-1.92& 3.32 $\times$ 1.47 at -3 & 0.035 &   ...  \\
\tableline
G041.22$-$00.19 &19:07:21.3720  & $+$07:17:08.100  & ...  & ... & 3.31 $\times$ 1.15 at -2 &  1.89 &$+$55.38 \\
J1905+0652      &19:05:21.2104  & $+$06:52:10.7803 &-0.50 &-0.42& 2.73 $\times$ 0.98 at  0 & 0.031 &   ...  \\
J1919+0619      &19:19:17.3502  & $+$06:19:42.7700 & 2.96 &-0.96& 3.22 $\times$ 0.85 at 16 & 0.158 &   ...  \\
J1856+0610      &18:56:31.8388  & $+$06:10:16.7650 &-2.68 &-1.11& 3.08 $\times$ 1.17 at  0 & 0.089 &   ...  \\
J1907+0907      &19:07:41.9634  & $+$09:07:12.397  & 0.09 & 1.83& 3.04 $\times$ 1.16 at  0 & 0.077 &   ...  \\
\tableline
G043.03$-$00.45 &19:11:38.99000  & $+$08:46:30.8000 & ...  & ... & 3.31 $\times$ 1.15 at -2 &  1.55 &$+$54.82 \\
J1907+0907      &19:07:41.96340  & $+$09:07:12.3970 &-0.98 & 0.34& 2.85 $\times$ 1.17 at -1 &  0.073&   ...  \\
J1913+0932      &19:13:24.02539  & $+$09:32:45.3791 & 0.43 & 0.77& 2.94 $\times$ 1.12 at -3 &  0.008&   ...  \\
J1905+0952      &19:05:39.89890  & $+$09:52:08.4070 &-1.48 & 1.09& 2.86 $\times$ 1.18 at  0 &  0.063&   ...  \\
J1922+0841      &19:22:18.63370  & $+$08:41:57.3730 & 2.63 &-0.08& 3.09 $\times$ 1.23 at -4 &  0.041&   ...  \\
\tableline
G043.89$-$00.78 &19:14:26.39050  & $+$09:22:36.5660 & ...  & ... & 3.70 $\times$ 1.37 at -9 &  3.85 &$+$48.00 \\
J1913+0932      &19:13:24.02539  & $+$09:32:45.3791 &-0.26 & 0.17& 3.45 $\times$ 1.36 at -9 &  0.009&   ...  \\
J1907+0907      &19:07:41.96336  & $+$09:07:12.3968 &-1.66 &-0.26& 3.53 $\times$ 1.33 at -10&  0.096&   ...  \\
J1922+0841      &19:22:18.63360  & $+$08:41:57.3780 & 1.94 &-0.68& 3.55 $\times$ 1.44 at -8 &  0.054&   ...  \\
J1905+0952      &19:05:39.89889  & $+$09:52:08.4071 &-2.16 & 0.49& 3.35 $\times$ 1.37 at -8 &  0.075&   ...  \\
\tableline
G045.49$+$00.12 &19 14 11.35530  & $+$11:13:06.3700 & ...  & ... & 3.51 $\times$ 1.20 at -6 &  3.10 &$+$57.36 \\
J1908+1201      &19:08:40.32064  & $+$12:01:58.8609 &-1.35 & 0.81& 3.14 $\times$ 1.34 at -4 &  0.063&   ...  \\
J1913+1307      &19:13:14.00638  & $+$13:07:47.3307 &-0.23 & 1.91& 3.07 $\times$ 1.21 at -5 &  0.087&   ...  \\
J1905+0952      &19:05:39.89890  & $+$09:52:08.4070 &-2.09 &-1.35& 3.09 $\times$ 1.22 at -3 &  0.076&   ...  \\
J1925+1227      &19:25:40.81710  & $+$12:27:38.0870 & 2.82 & 1.24& 3.18 $\times$ 1.28 at -4 &  0.041&   ...  \\
\tableline
G045.80$-$00.35 &19 16 31.07930  & $+$11:16:11.9880 & ...  & ...& 3.55 $\times$ 1.34 at -3 &  2.23 &$+$60.00 \\
J1908+1201      &19:08:40.32064  & $+$12:01:58.8609 &-1.92 &0.76& 3.17 $\times$ 1.30 at -2 &  0.078&   ...  \\
J1913+0932      &19:13:24.02539  & $+$09:32:45.3791 &-0.76 &-1.7& 3.17 $\times$ 1.30 at -2 &  0.155&   ...  \\
J1925+1227      &19:25:40.81708  & $+$12:27:38.0856 & 2.25 &1.19& 3.17 $\times$ 1.30 at -2 &  0.113&   ...  \\
J1913+1307      &19:13:14.00638  & $+$13:07:47.3307 &-0.81 &1.86& 3.17 $\times$ 1.30 at -2 &  0.079&   ...  \\
\tableline
G049.34$+$00.41 &19:20:32.4500  & $+$+14:45:45.400  & ...  & ... & 6.39 $\times$ 1.28 at -19&  1.772&$+$67.60 \\
J1922+1504      &19:22:33.27280 & $+$+15:04:47.5370 & 0.49 & 0.32& 7.17 $\times$ 1.12 at -20&  0.021&   ...  \\
J1922+1530      &19:22:34.69918 & $+$+15:30:10.0314 & 0.49 & 0.74& 6.24 $\times$ 1.74 at -17&  0.080&   ...  \\
J1917+1405      &19:17:18.06392 & $+$+14:05:09.7704 &-0.78 &-0.68& 6.62 $\times$ 1.34 at -18&  0.048&   ...  \\
J1924+1540      &19:24:39.45574 & $+$+15:40:43.9405 & 1.00 & 0.92& 6.18 $\times$ 1.64 at -19&  0.257&   ...  \\
\tableline
G052.10$+$01.04 &19:23:37.3295  & $+$17:29:10.495  & ...  & ... & 0.93 $\times$ 0.54 at -13&  6.330&$-$32.30 \\
J1921+1625      &19:21:57.38244 & $+$16:25:01.9231 &-0.40 &-1.07& 1.01 $\times$ 0 58 at -9 &  0.041&   ...  \\
J1927+1847      &19:27:32.31207 & $+$18:47:07.9048 & 0.93 & 1.30& 0.97 $\times$ 0.57 at -11&  0.016&   ...  \\
J1928+1859      &19:28:42.71875 & $+$18:59:24.5627 & 1.21 & 1.50& 0.94 $\times$ 0.60 at -11&  0.032&   ...  \\
J1924+1540      &19:24:39.45590 & $+$15:40:43.9420 & 0.25 &-1.81& 1.01 $\times$ 0.57 at -10&  0.267&   ...  \\
\enddata

\tablecomments{The R.A.$_{sep}$ is $\Delta$R.A.cos(Dec). }

\end{deluxetable}

\end{document}